\documentclass[trackchanges]{aastex7}
\PassOptionsToPackage{figuresleft}{rotating}

\usepackage[utf8]{inputenc}  
\usepackage{indentfirst}
\usepackage{xcolor,graphicx}
\usepackage{multirow}
\usepackage{booktabs}
\usepackage{subfigure}

\graphicspath{{./}{figures/}}

\begin{document}
\title{A re-identification of six Candidate Gravitationally Lensed Gamma-Ray Bursts}

\author[orcid=0009-0004-8586-1367]{Yu Zhao}
\affiliation{College of Physics and Electronics information, Yunnan Normal University, Kunming 650500, People's Republic of China}
\email[show]{zhaoyu99612@163.com}  

\author[orcid=0000-0003-3846-0988]{Zhao-Yang Peng} 
\affiliation{College of Physics and Electronics information, Yunnan Normal University, Kunming 650500, People's Republic of China}
\email[show]{pengzhaoyang412@163.com}

\author[orcid=0000-0001-5681-6939]{Jia-Ming Chen}
\affiliation{Department of Astronomy, School of Physics and Astronomy, Key Laboratory of Astroparticle Physics of Yunnan Province, Yunnan University, Kunming 650091,China}
\email{chenjiaming5821@163.com}  

\author{Yue Yin}
\affiliation{Department of Physics, Liupanshui Normal College, Liupanshui 553004, People's Republic of China\\}
\email{564089310@qq.com}
\author{Ting Li}
\affiliation{State-owned Assets and Laboratory Management Office, Yunnan Normal University, Kunming 650500, People's Republic of China}
\email{369271054@qq.com}
 
\begin{abstract}
The gravitational lensing effect of gamma-ray bursts (GRBs) holds significant and diverse applications in the field of astronomy. Nevertheless, the identification of millilensing events in GRBs presents substantial challenges. We re-evaluate the gravitational lensing candidacy of six previously proposed GRBs (GRB 081122A, GRB 081126A, GRB 090717A, GRB 110517B, GRB 200716C, and GRB 210812A) using a comprehensive set of temporal and spectral diagnostics.  These include $\chi^2$ light-curve similarity tests, photon-count–based hardness ratio (HR$_\mathrm{count}$) comparisons, $T_{90}$ duration measurements, spectral lag, Norris pulse-shape fitting, and both time-resolved and time-integrated spectral analyses.  We propose an evaluation framework, any single test that reveals a statistically significant inconsistency between the two pulses is sufficient to reject the lensing hypothesis for that burst.  Although certain diagnostics, such as $T_{90}$ and parametric model fits, have known limitations, they are applied and interpreted in conjunction with the more robust, model-independent $\chi^2$ and HR$_\mathrm{count}$ tests.  For all six GRBs, at least one diagnostic shows a significant discrepancy, leading us to conclude that none are consistent with a gravitational lensing interpretation.
\end{abstract}

\section{Introduction\label{section1}}
Gravitational lensing occurs when a massive object lies near the line of sight between an observer and a light source. Depending on the relative positions of the source, lens, and observer, as well as the mass and shape of the lens, gravitational lensing can produce multiple images of a single source  \citep{1964MNRAS.128..295R,1979Natur.279..381W,1992grle.book.....S}. In the case of a point-mass gravitational lens, two distinct images of the source are formed, and the overall brightness of the source is magnified. While the intensities of these images differ, it is important to note that their spectral shapes remain identical.

Gamma-ray bursts (GRBs) are one of the brightest and most abundant high-energy transient phenomena observed since their discovery by the US Vela satellite in 1967, and \cite{1973ApJ...182L..85K} were the first to publish a scientific account of GRBs. With the launch and successful operation of several dedicated detectors, such as the Burst and Transient Source Experiment (BATSE) at the Compton Gamma-ray Observatory (CGRO; \citep{1992Natur.355..143M,1999ApJS..122..465P}), the Burst Alert Telescope (BAT) at the Neil Gehrels Swift Observatory \citep{2004ApJ...611.1005G,2005SSRv..120..143B}, and the Gamma-ray Burst Monitor (GBM) at the Fermi Observatory \citep{2009ApJ...702..791M}, we have gained a better understanding of GRBs. For example, spectroscopic measurements of the redshift of the host galaxy indicate that GRBa occur at cosmological distances \citep{1997Natur.387..878M}, and their spectroscopic measurements range from 0.0085 \citep{1998Natur.395..670G, 1998Natur.395..663K} to 9.4 \citep{2011ApJ...736....7C}. Since GRBs have a large range(up to redshift z$<{9}$; \citep{2011ApJ...736....7C}) of distances, this means that some of the radiation from GRBs may be lensed by foreground objects before reaching us \citep{1986ApJ...308L..43P,1992ApJ...389L..41M}. In such scenarios, compact foreground objects—such as primordial black holes, neutron stars, or other dark compact remnants—can act as gravitational lenses, amplifying or duplicating GRB signals and introducing time delays on the order of milliseconds to seconds. Detecting gravitational lensing signatures in GRBs provides a valuable opportunity to probe the existence and distribution of compact objects in the universe, especially those that are electromagnetically dark. While strong gravitational lensing has been extensively studied in the optical and radio bands for distant galaxies and quasars, identifying lensing features in GRB light curves remains challenging due to their highly transient nature and short durations. Nevertheless, with high time-resolution observations and detailed spectral analyses, it is possible to search for potential lensing indicators, such as repeated pulses with similar temporal structures and spectral characteristics.

The application of gravitational lensing to GRB can be roughly divided into two cases. One is that an event triggers the same instrument twice, and the two triggers will have similar light curve shapes and spectra, also known as macrolensing event \citep{1964MNRAS.128..307R,1994ApJ...432..478N,2009arXiv0912.3928V,2011AIPC.1358...17D,2014SCPMA..57.1592L,2019ApJ...871..121H,2020ApJ...897..178A,2021NatAs...5.1118R}. Its lens candidates are usually galaxies or galaxy clusters (mass$> {10}^{7}\,{M}_{\odot}$). After observing approximately 10,000 GRBs over three decades, no convincing large-lens candidate GRB pairs have been found \citep{1994ApJ...432..478N,2009arXiv0912.3928V,2011AIPC.1358...17D,2019ApJ...871..121H,2020ApJ...897..178A}. Another is that there are two emission events with similar light curve patterns and similar spectral characteristics in a single triggered GRB, and the signature of gravitational lensing is imprinted on the light curve of a single trigger, also known as millilensing event \citep{2001PhRvL..86..580N}. Such lenses produce angular separations between images on the order of milliarcseconds and corresponding time delays ranging from milliseconds to minutes. The lens candidate could be an intermediate-mass black hole \citep{2021NatAs...5..560P} with a mass of ${10}^{4}-{10}^{7}\,{M}_{\odot}$. This is distinct from microlensing, which typically refers to lensing by stellar-mass objects (angular separations of microarcseconds and delays of microseconds to milliseconds). In the context of GRBs, a possible millilensing could be an intermediate-mass black hole \citep{2021NatAs...5..560P}. Theoretical estimates indicate that millilensing of GRBs is expected to be rare under standard cosmological and compact-object assumptions. \cite{2001PhRvL..86..580N} showed that optimistic scenarios (with a very large fraction of dark matter in $\sim10^{7}M_\odot$ compact objects) yield at most of order 10\% of GRBs being millilensed, more realistic scenarios predict a much lower rate.

However, several candidates for millilensing events have recently been proposed. \cite{2021NatAs...5..560P} discovered evidence of gravitational millilensing in the light curve of GRB 950830 and inferred that the lensing object was an intermediate-mass black hole. However, \cite{2021RNAAS...5..103M} identified a cumulative hardness difference between the two pulses of GRB 950830, suggesting that while the case may be intriguing in the context of gravitational lensing, it should not be considered confirmed. Subsequently, \cite{2021ApJ...922...77K} conducted studies suggesting the presence of millilensing effects in GRB 090717A. However, \cite{2021RNAAS...5..183M} argued that the light curves of the two pulses in GRB 090717A differ at a $5\sigma$ confidence level, and therefore concluded that GRB 090717A does not represent a compelling case of gravitational lensing. Notably, these studies focused primarily on light-curve morphology and hardness ratios, without a systematic analysis of time-resolved or time-integrated spectra. Shortly after, papers suggesting the presence of millilensing in GRB 200716C were published by \cite{2021ApJ...921L..29Y} and \cite{2021ApJ...918L..34W}. They conducted a detailed spectral analysis of GRB 200716C and found evidence supporting its classification as a gravitationally lensed GRB. However, \cite{2022RNAAS...6...42M} performed a $\chi^2$ test on the two pulses of GRB 200716C, which indicated that the probability of the two pulses being images of the same parent pulse is less than 0.218\%, differing at about the $3.1\sigma$ confidence level. Therefore, GRB 200716C is not considered a compelling case for gravitational lensing. Apart from these three GRBs, four other GRBs have been proposed as strong candidates for millilensing. \cite{2021ApJ...921L..30V} conducted a detailed temporal and spectral analysis and claimed that GRB 210812A is gravitationally lensed. Another study by \cite{2022ApJ...931....4L} identified four potential cases—GRB 081126A, GRB 090717A, GRB 081122A, and GRB 110517B—with varying levels of confidence. However, \cite{2024MNRAS.529L..83M} conducted light curve and hardness similarity tests on these GRBs and found that except for GRB 210812A, none of them showed clear evidence of millilensing.

Based on the above studies, a consensus remains elusive regarding whether these GRBs are indeed gravitationally lensed events, necessitating more compelling evidence to validate such claims. To address this, our study aims to conduct a detailed analysis to determine whether these bursts represent genuine cases of gravitational lensing. Specifically, we seek to systematically re-identify these candidate GRB gravitational lensing events. Our approach involves not only comparing their spectral components and features using optimal fit models but also leveraging temporal characteristics to provide evidence for evaluating their nature as gravitationally lensed GRBs. In this work we adopt a conservative evaluation criterion: if any independent, robust test conclusively rejects a lensing interpretation for a candidate, that candidate is excluded as a gravitational millilensing event. We therefore require consistency across all applied temporal and spectral tests before regarding a case as a viable lens candidate. The paper is organized as follows. Section \ref{section 2} details the criteria for data selection and the methodology used in this study. The models used in this study and the criteria for selecting the best model are presented in Section \ref{section 3}. Section \ref{section 4} systematically presents the lens analysis and its main findings. Section \ref{section 5} provides the discussion and conclusions. 

\section{Sample Selection and Analysis Methods\label{section 2} }
The GRB sample utilized in this study was acquired from the GBM aboard the Fermi Gamma-ray Space Telescope. The GBM comprises 14 detector modules, including 12 sodium iodide (NaI) detectors and 2 bismuth germanate (BGO) detectors. The NaI detectors cover an energy range of 8 keV to 1 MeV, while the BGO detectors extend from 200 keV to 40 MeV. GBM data are available in three formats: CTIME, CSPEC, and Time-Tagged Event (TTE) files, which were downloaded from the Fermi Science Support Center (FSSC) FTP site. For spectral and temporal analysis, we employ TTE data with a $2~\mu\mathrm{s}$ time resolution, spanning a short time window of approximately $-30$ to $300$ seconds. Data were selected from all NaI detectors triggered by the GBM (typically detectors 0–3) and the brightest BGO detector. To facilitate comprehensive analysis and comparison, both time-integrated and time-resolved spectral analyses were performed for each GRB in the sample.

In this work, the Multi-Mission Maximum Likelihood Framework (3ML; \cite{2015arXiv150708343V}) package serves as the primary tool for the spectral analysis, which widely adopted in GRB spectroscopy \citep{2018ApJ...857..120Y}. The method has been successfully applied in prior studies \citep{2021ApJ...920...53C,2022ApJ...932...25C,2024ApJ...970...67D,2024ApJ...969...26P}. For time-resolved spectral analysis, light curves must be partitioned into sufficiently fine intervals. \cite{2014MNRAS.445.2589B} demonstrated that the Bayesian block algorithm yields optimally refined time bins, though this method does not ensure each bin contains adequate photons for accurate spectral fitting. To address this, we first apply the Bayesian block algorithm to segment the data into time bins and then calculate the statistical significance S (signal-to-noise ratio) for each bin. Given the relatively low peak fluxes of gravitationally lensed GRBs, only bins with $S>5$  were selected for spectral analysis. Additionally, the Bayesian block method with a false positive rate of $p=0.01$ was used to reconstruct the TTE light curve of the brightest NaI detector, with the same block structure applied to the remaining detectors. 

\section{Spectral Models and Selection of the Best Model\label{section 3}}
For each spectral fit, three distinct models were employed: the cutoff power-law (CPL), Band function, and blackbody function (BB). The Band spectral component, described by the Band function \citep{1993ApJ...413..281B}, is expressed as
\begin{equation}
{N_{B{\rm{and}}}}\left( E \right) = A\left\{ \begin{array}{l}
\left( {\frac{E}{{100keV}}} \right)\exp \left( { - \frac{E}{{{E_0}}}} \right),E < \left( {\alpha  - \beta } \right){E_0}\\
{\left( {\frac{{\left( {\alpha  - \beta } \right){E_0}}}{{100keV}}} \right)^{\alpha  - \beta }}\exp \left( {\beta  - \alpha } \right){\left( {\frac{E}{{100keV}}} \right)^\beta },E > \left( {\alpha  - \beta } \right){E_0}
\end{array} \right.
\end{equation}
where
\begin{equation}
{E_p} = \left( {2 + \alpha } \right){E_0},
\end{equation}
Here, $N_{E}$ represents the photon flux (ph $cm^{-2} keV^{-1} s^{-1}$), where $A$ is the normalization constant, $\alpha$ and $\beta$ are the low- and high-energy spectral indices, respectively. The parameter ${E_0}$ is the break energy, while $E_{p}$ indicates the peak energy (in keV) of the observed $\nu {F_\nu }$ spectrum. 

The cutoff power-law function (CPL) \citep{2011MNRAS.411.1323G} is expressed as
\begin{equation}
{N_{CPL}}\left( E \right) = A{\left( {\frac{E}{{100keV}}} \right)^\alpha }\exp \left( { - \frac{E}{{{E_0}}}} \right),
\end{equation}
where $A$ is the normalization factor at 100 keV, given in units of photons ${s^{ - 1}}c{m^{ - 2}}ke{V^{ - 1}}$, $\alpha$ is the low-energy spectral index, and ${E_0}$ denotes the break energy in keV. 

Certain GRBs exhibit additional thermal components, which are typically modeled using the Planck blackbody (BB) function. The Planck function is expressed as:
\begin{equation}
N_{B B}(E)=A \frac{E^2}{\exp [E / k T]-1},
\end{equation}
where $E$ denotes the photon energy, $A$ is the normalization constant defined at 1 keV, $kT$ represents the blackbody temperature in units of keV, and $k$ is the Boltzmann constant.

In our analysis, we first employ the Band and CPL function models to fit the time slices of the two pulses of GRB. We then select the optimal fitting results to determine the best-fit model. Subsequently, we incorporate the BB component into the Band and CPL function models, respectively, yielding Band+BB and CPL+BB models. Through data analysis, we again identify the best model containing the BB component (best model+BB). Finally, we compare the best model and the best model+BB to determine the existence of the BB component.

To assess the quality of the fitting models, we adopt the Deviance Information Criterion (DIC; \cite{2019ApJS..245....7L,2021ApJ...920...53C}). The DIC is expressed as:
\begin{equation}
DIC = -2 \log \left[ p(\text{data} \mid \hat{\theta}) \right] + 2 p_{\text{DIC}},
\end{equation}
Here, $\hat{\theta}$ denotes the posterior mean of the parameters, and $p_{\text{DIC}}$ represents the effective number of parameters. Multiple models are fitted to the same dataset, and a lower DIC value indicates a better model fit. The difference in DIC values between two models, defined as $\Delta \text{DIC} = \text{DIC}_j - \text{DIC}_i$, is used to evaluate their relative performance. According to the criterion outlined by \citet{2018ApJ...866...13H}:
(a) $\Delta\text{DIC} = 0-2$: the models are indistinguishable in terms of fit quality;
(b) $\Delta\text{DIC} = 2-6$: there is positive evidence favoring model $i$;
(c) $\Delta\text{DIC} = 6-10$: there is strong evidence in support of model $i$;
(d) $\Delta\text{DIC} > 10$: there is very strong evidence favoring model $i$.

\section{Analysis and Results of Gravitational Lensing\label{section 4}}
\subsection{Temporal Characteristics}
\subsubsection{Duration}
We compute the $T_{90}$ duration (8-1000 keV, defined as the time interval during which the detector accumulates from 5\% to 95\% of the total fluence) for each pulse of all GRBs, with the results summarized in Table \ref{table 1}. Taking GRB 200716C as an example for analysis, its two pulses exhibit consistent durations: the first pulse has a $T_{90,1} = 0.71 \pm 0.01 \, \text{s}$, while the second pulse has a $T_{90,2} = 0.73 \pm 0.03 \, \text{s}$. By treating the first pulse as an independent GRB and classifying it according to its $T_{90}$ distribution \citep{2016ApJS..223...28N,2020ApJ...893...46V}, we categorize these events as either long GRB (likely originating from the collapse of massive stars) or short GRB (potentially resulting from compact binary mergers). For GRB 081122A, the duration of the first pulse is $T_{90,1} = 4.87 \pm 0.13 \, \text{s} $, and the duration of the second pulse is  $T_{90,2} = 6.67 \pm 0.11 \, \text{s} $. The difference in duration between the two pulses is far greater than the $3\sigma $ confidence level. The $T_{90}$ durations of the remaining GRBs differ, with the exception of GRB 200716C, which is consistent within $1\sigma$, while all others differ at $>3\sigma$. Furthermore, we find that, with the exception of GRB 200716C, which is a short GRB, the remaining bursts are all long GRBs.
\subsubsection{Spectral Lag}
In the prompt emission phase of GRBs, the observed temporal delay between soft and hard photons (typically in the range of several keV to hundreds of keV) is referred to as spectral lag \citep{1995A&A...300..746C}. A GRB exhibits positive spectral lag when higher-energy photons arrive earlier than lower-energy photons, while the reverse case indicates negative spectral lag. We have calculated the spectral lags for different pulses across various GRBs using cross-correlation function (CCF) method \citep{1997ApJ...486..928B}. In this work, we adopt an improved CCF technique \citep{2015MNRAS.446.1129B} that accounts for potential asymmetry in the CCF by fitting an asymmetric Gaussian model. Our analysis focuses on two energy bands: 15–25 keV and 50–100 keV. The derived spectral lag parameters are presented in Table \ref{table 1}. Taking GRB 081122A as an example, the two pulses exhibit the same lag: the first pulse has $\tau_{\text{lag},1} = -0.12 \pm 0.04 \, \text{s}$, while the second pulse has $\tau_{\text{lag},2} = -0.12 \pm 0.04 \, \text{s}$. In contrast, For GRB 200716C, the first pulse has $\tau_{\text{lag},1} = -0.02 \pm 0.01 \, \text{s}$, and the second pulse has $\tau_{\text{lag},2} = -0.12 \pm 0.04 \, \text{s}$, showing significant differences in lag between the two pulses. The same analysis is also conducted for the remaining bursts. The study finds that, except for GRB 200716C, the lags of the two pulses within other bursts are consistent within the $1\sigma$ region. Additionally, it is observed that all bursts exhibit negative lags except GRB 210812A, which shows a positive lag.

\begin{table}[htbp]
\centering
\caption{Temporal Characteristics, HR from spectral fits, and HR from photon counts of GRBs}
\label{table 1}
\resizebox{\textwidth}{!}{%
\begin{tabular}{lcccccccccc}
\hline
GRB Name  & \multicolumn{2}{c}{Time Period (s)} & \multicolumn{2}{c}{$T_{90}$ (s)} 
& \multicolumn{2}{c}{HR (fit)} & \multicolumn{2}{c}{HR$_\mathrm{count}$} & \multicolumn{2}{c}{Lag (s)} \\
\cline{2-11}
 & 1st & 2nd & 1st & 2nd & 1st & 2nd & 1st & 2nd & 1st & 2nd \\
\hline
GRB 081122A  & -3,6  & 10,19  & $4.87 \pm 0.13$  & $6.67 \pm 0.11$ 
& $5.76 \pm 1.23$  & $3.40 \pm 1.50$ & $1.23 \pm 0.09$ & $0.90 \pm 0.13$
& $-0.12 \pm 0.04$ & $-0.12 \pm 0.04$ \\
GRB 081126A  & -4,8  & 28,40  & $8.08 \pm 0.15$  & $8.81 \pm 0.14$  
& $5.52 \pm 0.83$  & $5.05 \pm 0.98$ & $1.27 \pm 0.13$ & $1.19 \pm 0.16$
& $-0.93 \pm 0.07$ & $-0.99 \pm 0.01$ \\
GRB 090717A  & -3,25  & 40,68  & $19.53 \pm 0.20$ & $20.37 \pm 0.07$
& $2.50 \pm 0.41$ & $2.82\pm 0.57$ & $0.76 \pm 0.04$ & $0.86 \pm 0.06$
& $-0.99 \pm 0.01$ & $-0.96 \pm 0.06$ \\
GRB 110517B  & -3,8   & 15,26  & $6.84 \pm 0.09$ & $7.98 \pm 0.10$
& $4.20 \pm 0.74$ & $4.55 \pm 0.80$ & $0.93 \pm 0.07$ & $0.88 \pm 0.07$
& $-0.99 \pm 0.02$ & $-0.95 \pm 0.05$ \\
GRB 200716C  & 0,1    & 2,3    & $0.71 \pm 0.01$ & $0.73 \pm 0.03$  
& $5.29 \pm 0.50$ & $5.55 \pm 0.47$ & $1.62 \pm 0.11$ & $1.38 \pm 0.12$
& $-0.02 \pm 0.01$ & $-0.12 \pm 0.01$ \\
GRB 210812A  & -2,7   & 32,41  & $6.19 \pm 0.10$ & $7.68 \pm 0.08$  
& $3.74 \pm 0.84$ & $1.92 \pm 0.53$ & $1.41 \pm 0.20$ & $0.67 \pm 0.19$
& $0.19 \pm 0.12$ & $0.18 \pm 0.06$ \\
\hline
\end{tabular}%
}
\end{table}

\subsubsection{Light Curve Properties}
\begin{figure}[htbp]
\centering
\subfigure[Pulse 1 of GRB 081122A]{
\includegraphics[width=0.23\textwidth]{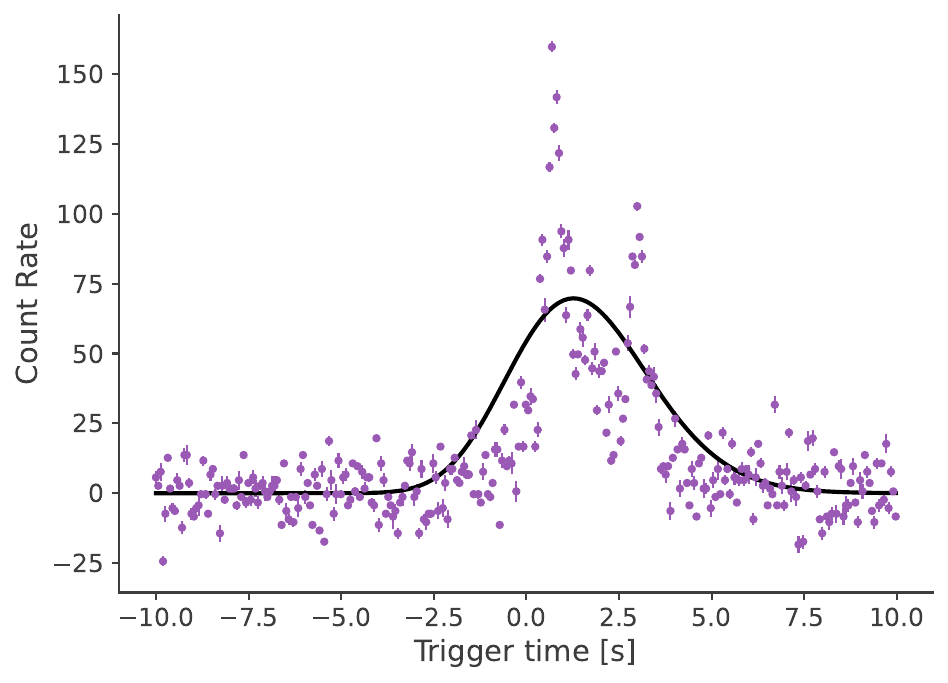 }
}
\subfigure[Pulse 2 of GRB 081122A]{
\includegraphics[width=0.23\textwidth]{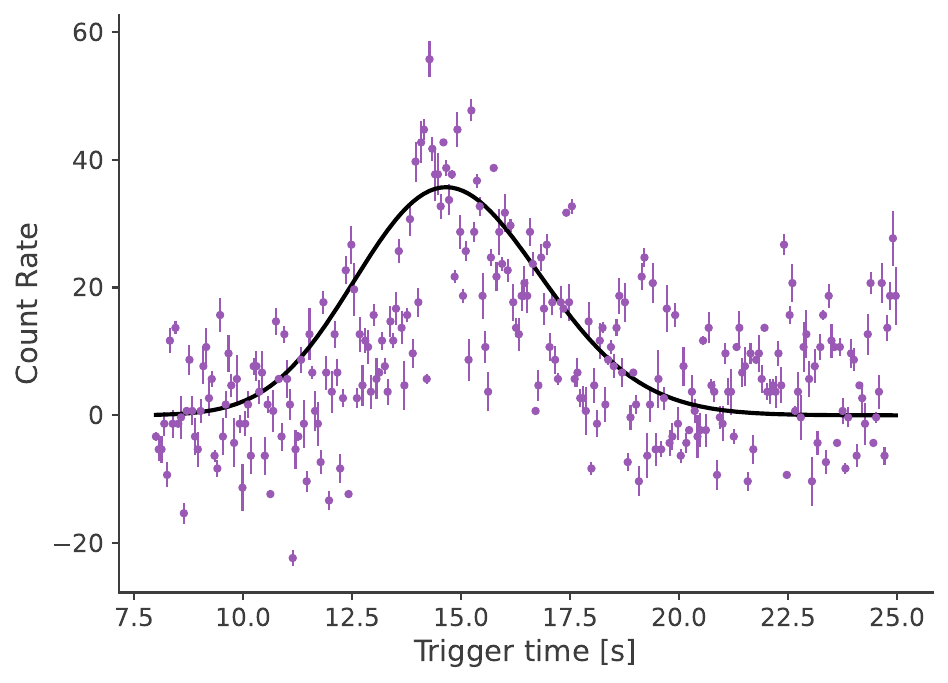}
}
\subfigure[Pulse 1 of GRB 081126A]{
\includegraphics[width=0.23\textwidth]{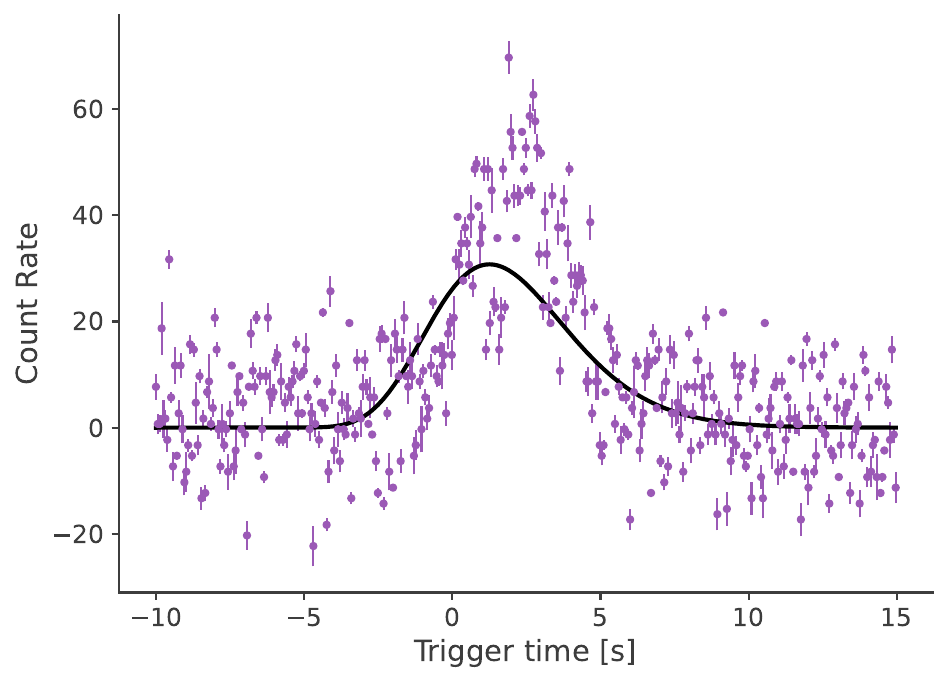}
}
\subfigure[Pulse 2 of GRB 081126A]{
\includegraphics[width=0.23\textwidth]{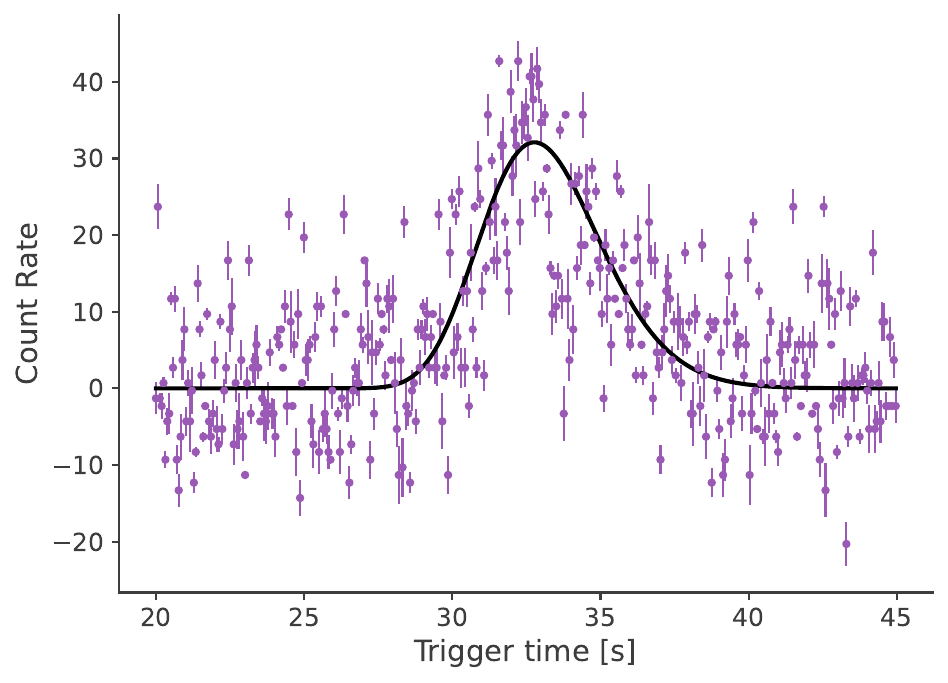}
}
\\
\subfigure[Pulse 1 of GRB 090717A]{
\includegraphics[width=0.23\textwidth]{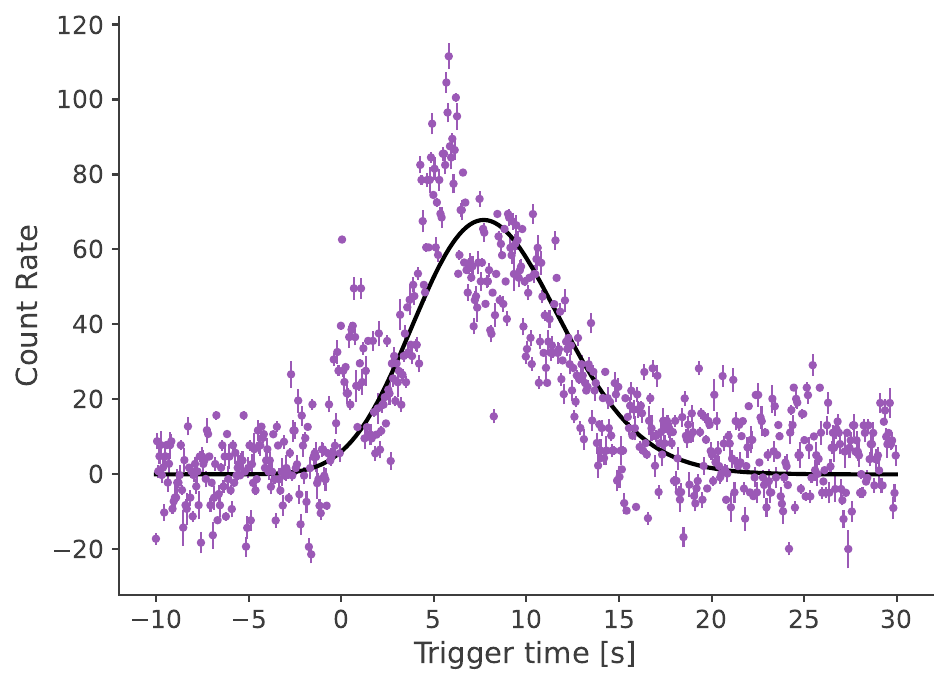}
}
\subfigure[Pulse 2 of GRB 090717A]{
\includegraphics[width=0.23\textwidth]{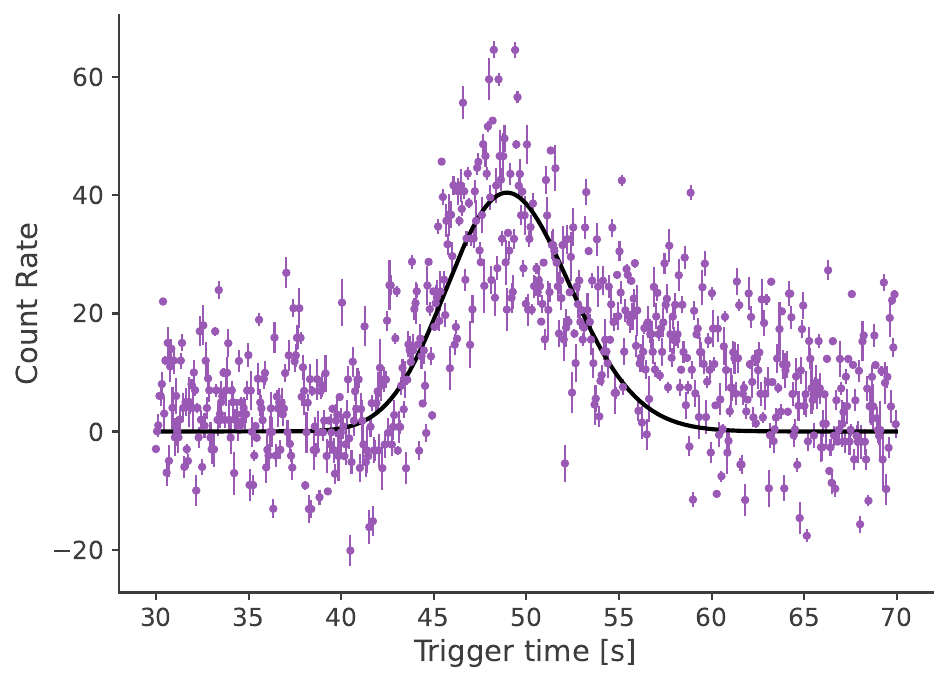}
}
\subfigure[Pulse 1 of GRB 110517B]{
\includegraphics[width=0.23\textwidth]{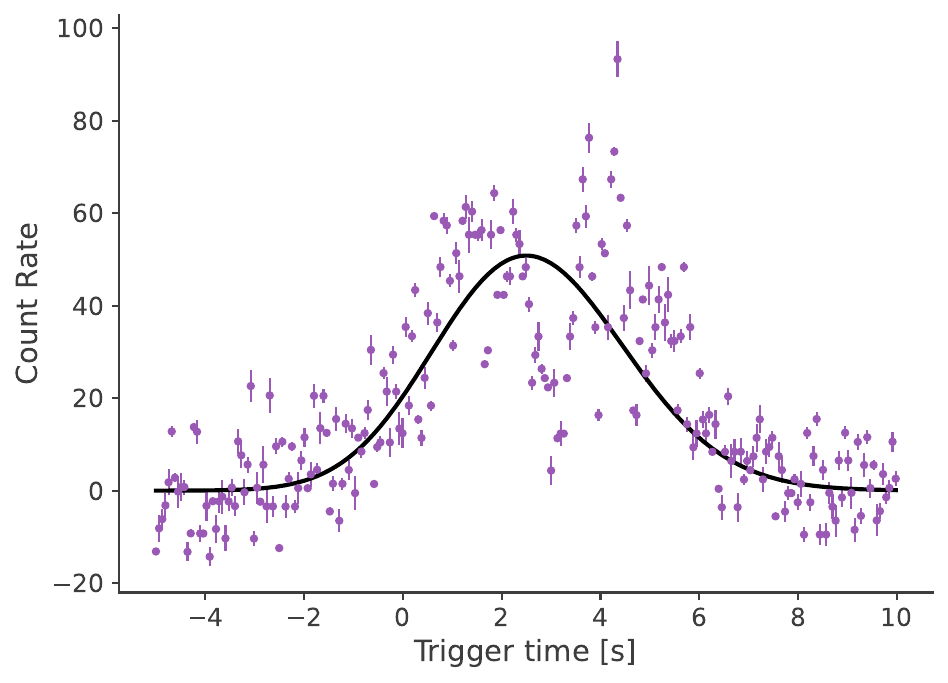}
}
\subfigure[Pulse 2 of GRB 110517B]{
\includegraphics[width=0.23\textwidth]{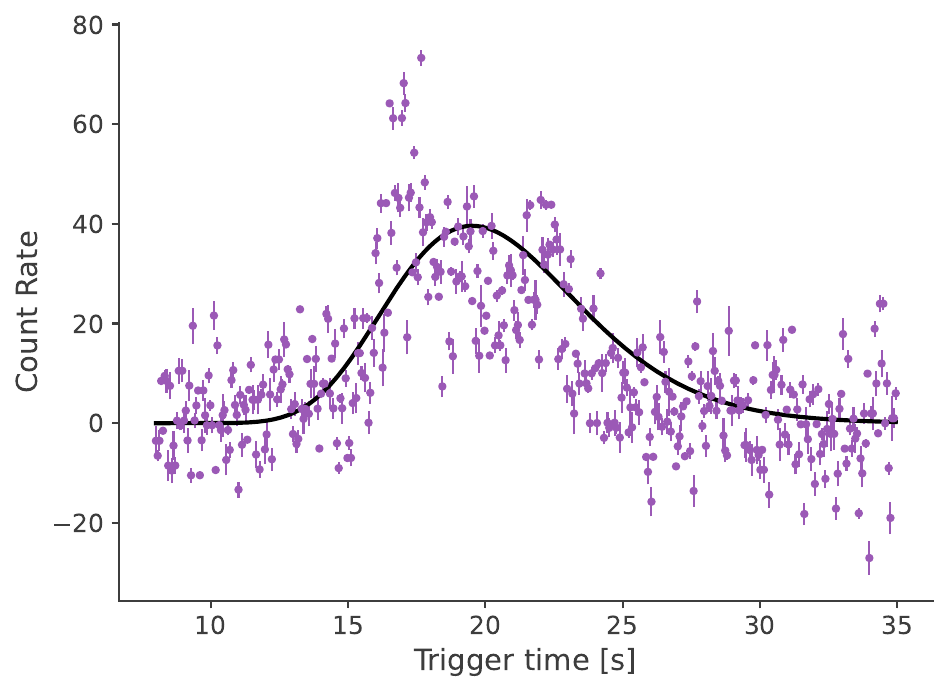}
}
\\
\subfigure[Pulse 1 of GRB 200716C]{
\includegraphics[width=0.23\textwidth]{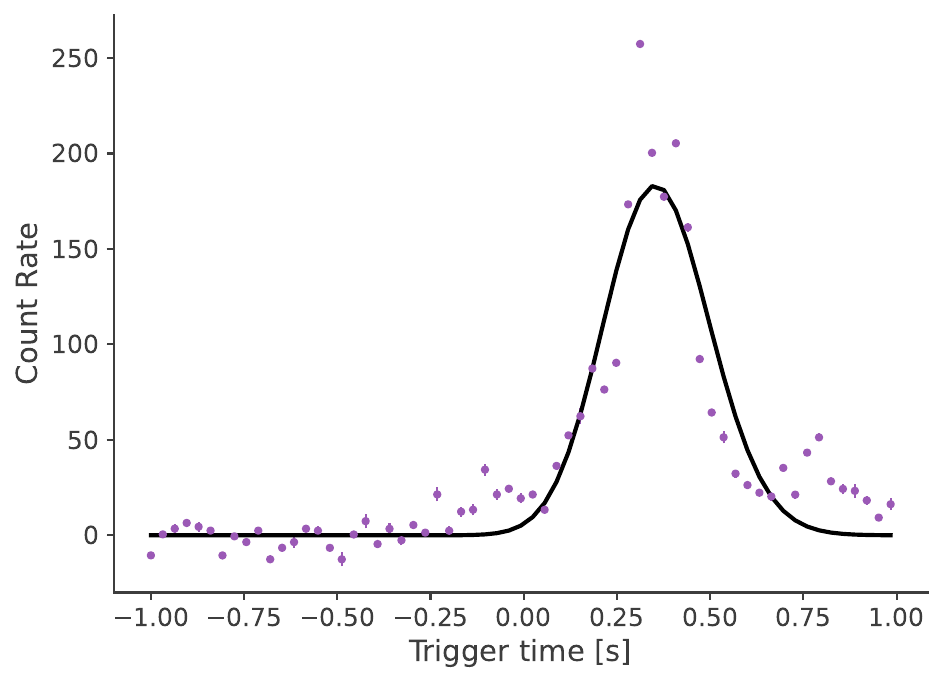}
}
\subfigure[Pulse 2 of GRB 200716C]{
\includegraphics[width=0.23\textwidth]{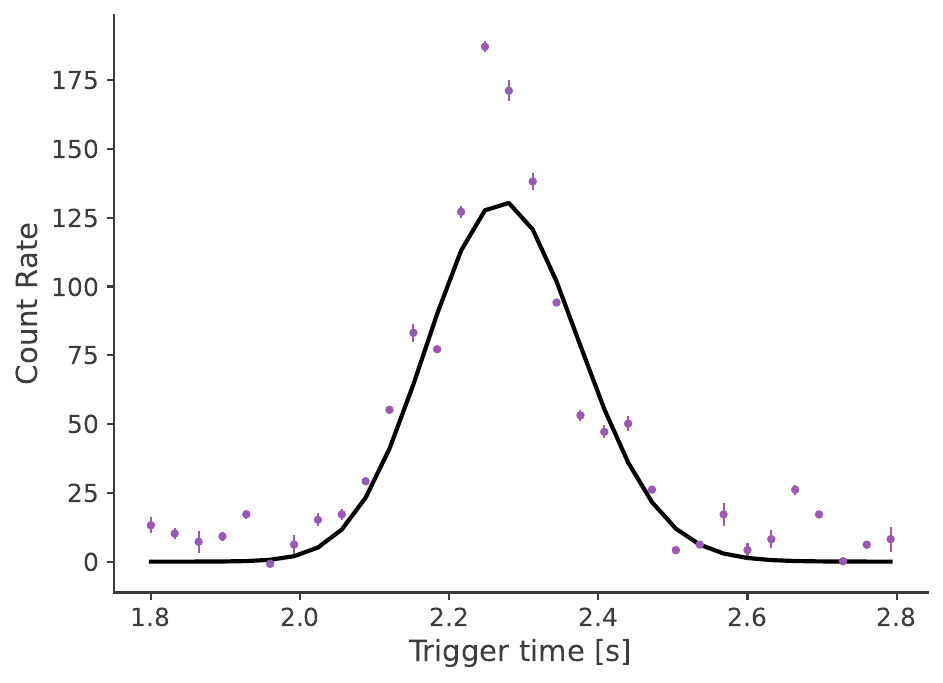}
}
\subfigure[Pulse 1 of GRB 210812A]{
\includegraphics[width=0.23\textwidth]{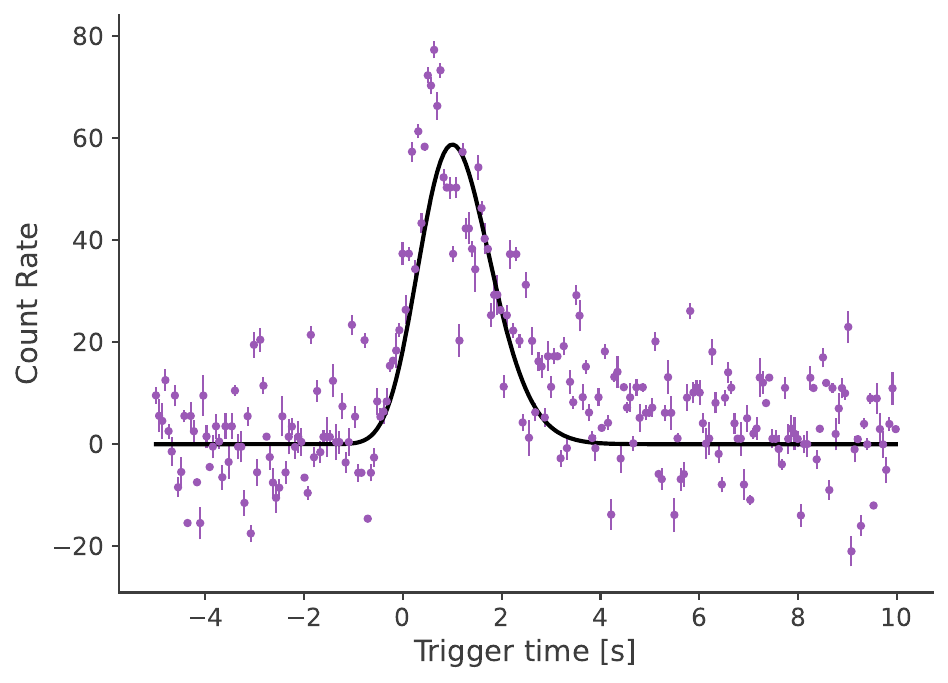}
}
\subfigure[Pulse 2 of GRB 210812A]{
\includegraphics[width=0.23\textwidth]{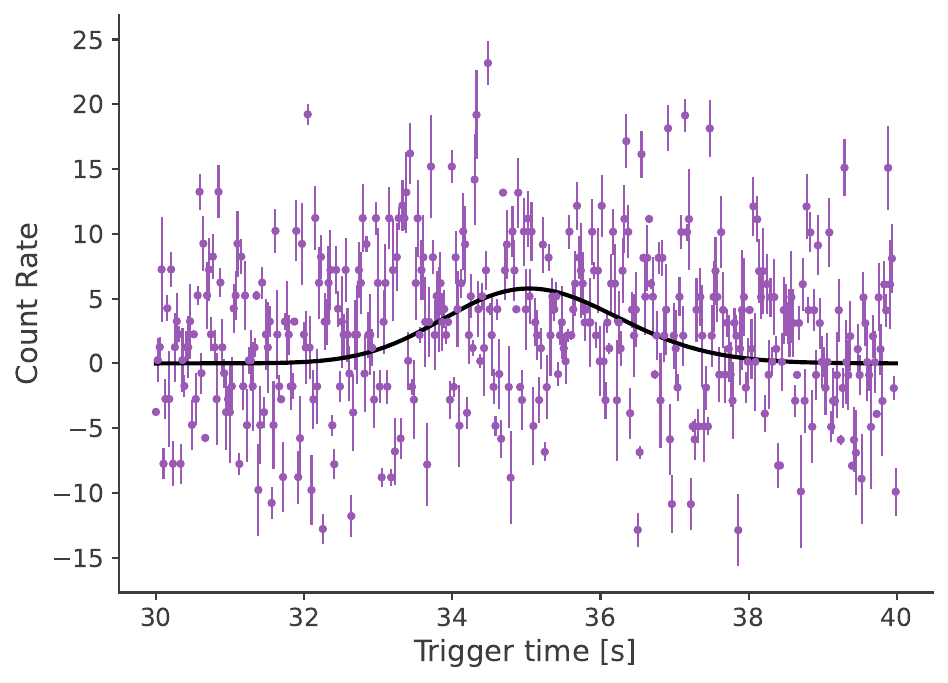}
}
\caption{Norris function fitting results of all GRB pulses.}
\label{fig:12pulses}
\end{figure}

To investigate the temporal properties of the pulses, we model the GRB pulses using the empirical pulse function proposed by \cite{2005ApJ...627..324N}.
\begin{equation}
I\left(t\right)=A\lambda\exp\left[-\tau_1/\left(t-t_s\right)-\left(t-t_s\right)/\tau_2\right]
\end{equation}
where $t$ denotes the time since the trigger, $A$ is the pulse amplitude, $t_s$ is the onset time of the pulse, and $\tau_1$ and $\tau_2$ characterize the rise and decay times of the pulse, respectively. The constant $\lambda$ is defined as $\lambda = \exp\left[2\left(\tau_1/\tau_2\right)^{1/2}\right]$.
We select the energy range of 8 keV to 1 MeV for the fitting, and the fitting results are shown in Figure~\ref{fig:12pulses}. Based on the fitting parameters, we obtain the temporal characteristics of the pulses (see Table~\ref{tab:parameter_summary}, including the pulse width $w=\tau_2{(1+2\ln{\lambda})}^{1/2}$, the pulse rise width, $t_{rise}=\frac{1}{2}w(1-k)$, and the decay width, $t_{decay}=\frac{1}{2}w(1+k)$. The pulse peak times are given by $t_{peak}=t_s+{(\tau_1/\tau_2)}^{1/2}$.
We take GRB081122A as an example. The first pulse has a width of $ w_1 = 5.35 \pm 0.31 \, \text{s} $, a rise time of $ t_{r,1} = 2.46 \pm 0.14 \, \text{s} $, and a decay time of $ t_{d,1} = 2.89 \pm 0.17 \, \text{s} $. The second pulse has a width of $ w_2 = 6.00 \pm 0.05 \, \text{s} $, a rise time of $ t_{r,2} = 2.87 \pm 0.03 \, \text{s} $, and a decay time of $ t_{d,2} = 3.13 \pm 0.03 \, \text{s} $. We find that only the decay time ($t_d$) is consistent within the 2$\sigma$ confidence level, while both the pulse width ($w$) and the rise time ($t_r$) show significant inconsistencies. The same analysis applies to the remaining bursts. We find that the pulse widths, rise times, and decay times of the two pulses in the remaining bursts all exhibit inconsistencies.
 
Additionally, we perform a $\chi^2$ test to examine the similarity of the light curves between two events. This test treats the binned light curves of the two pulses as two distinct distributions and evaluates their consistency with originating from the same parent distribution \citep{2021RNAAS...5..183M}. We conduct exploratory tests over a wide range of time-bin sizes to identify the settings that yield the most reliable $\chi^2$ results for each burst. 
For each GRB, we adopt a bin size that preserves sufficient pulse-structure detail while avoiding a regime where statistical fluctuations dominate the pulse profile. 
Furthermore, for each pulse, we select the fitting interval based on both the morphology of the light curve and photon-counting statistics. 
The chosen interval covers the full main emission episode while minimizing the influence of any trailing or preceding background. 
These fitting windows are determined through visual inspection combined with quantitative count-rate thresholds, and are applied consistently across all GRBs to enable symmetric and meaningful pulse-shape comparisons. Figure~\ref{fig2} shows the light curves of each GRB with the corresponding time bins.
The results are as follows: GRB 081122A ( 0.128\,s) yields $\chi^2 = 118.39$ ($p = 0.0002$); GRB 081126A ( 0.128\,s) has $\chi^2 = 121.98$ ($p = 0.0199$); GRB 090717A ( 0.128\,s) shows $\chi^2 = 265.66$ ($p = 0.0135$); GRB 110517B (0.128\,s) exhibits $\chi^2 = 283.32$ ($p = 0.00001$); GRB 200716C (0.008\,s) returns $\chi^2 = 292.83$ ($p = 0.00001$); and GRB 210812A ( 0.256\,s) has $\chi^2 = 31.66$ ($p = 0.5828$). GRB 210812A shows no significant deviation from the hypothesis that their pulse light curves are drawn from the same parent distribution ($p > 0.05$). In contrast, GRBs 081122A, 081126A, 090717A ,110517B, and GRB 200716C exhibit significant inconsistencies ($p < 0.05$).

\begin{longtable}{l *{8}{c}}
\caption{Parameter Summary}
\label{tab:parameter_summary}\\
\toprule
GRB & $A(\text{count}~\text{s}^{-1}
)$ & $\tau_1(\text{s})$ & $\tau_2(\text{s})$ & $t_s(\text{s})$ & $t_p(\text{s})$ & $w(\text{s})$ & $t_r(\text{s})$ & $t_d(\text{s})$ \\
\midrule

081122A(1) & $69.74 \pm 4.08$ & $626.34 \pm 17.06$ & $0.43 \pm 0.03$ & $-15.17 \pm 0.03$ & $1.27 \pm 0.68$ & $5.35 \pm 0.31$ & $2.46 \pm 0.14$ & $2.89 \pm 0.17$ \\
081122A(2) & $35.68 \pm 0.15$ & $4516.76 \pm 115.19$ & $0.26 \pm 0.01$ & $-19.69 \pm 0.01$ & $14.65 \pm 0.46$ & $6.00 \pm 0.05$ & $2.87 \pm 0.03$ & $3.13 \pm 0.03$ \\
081126A(1) & $30.76 \pm 0.15$ & $132.35 \pm 1.15$ & $0.96 \pm 0.01$ & $-10.01 \pm 0.01$ & $1.25 \pm 0.06$ & $6.64 \pm 0.03$ & $2.84 \pm 0.01$ & $3.80 \pm 0.02$ \\
081126A(2) & $32.11 \pm 0.16$ & $292.79 \pm 10.62$ & $0.61 \pm 0.01$ & $19.44 \pm 0.01$ & $32.77 \pm 0.26$ & $5.72 \pm 0.08$ & $2.56 \pm 0.03$ & $3.16 \pm 0.04$ \\
090717A(1) & $67.87 \pm 0.03$ & $1741.01 \pm 126.89$ & $0.82 \pm 0.02$ & $-30.03 \pm 0.02$ & $7.70 \pm 1.45$ & $11.14 \pm 0.29$ & $5.16 \pm 0.14$ & $5.98 \pm 0.15$ \\
090717A(2) & $40.47 \pm 0.15$ & $7946.97 \pm 675.27$ & $0.40 \pm 0.01$ & $-7.09 \pm 0.01$ & $48.95 \pm 2.48$ & $9.42 \pm 0.27$ & $4.51 \pm 0.13$ & $4.91 \pm 0.14$ \\
110517B(1) & $50.85 \pm 0.15$ & $4722.14 \pm 279.38$ & $0.23 \pm 0.01$ & $-30.21 \pm 0.01$ & $2.50 \pm 1.04$ & $5.45 \pm 0.12$ & $2.61 \pm 0.06$ & $2.84 \pm 0.06$ \\
110517B(2) & $35.07 \pm 0.13$ & $7705.28 \pm 535.61$ & $0.40 \pm 0.01$ & $-35.57 \pm 0.01$ & $20.02 \pm 2.04$ & $9.45 \pm 0.24$ & $4.52 \pm 0.11$ & $4.93 \pm 0.12$ \\
200716C(1) & $183.26 \pm 0.46$ & $1086.13 \pm 13.91$ & $0.01 \pm 0.0001$ & $-3.17 \pm 0.0001$ & $0.35 \pm 0.02$ & $0.40 \pm 0.001$ & $0.20 \pm 0.01$ & $0.21 \pm 0.01$ \\
200716C(2) & $130.94 \pm 0.94$ & $750.49 \pm 25.33$ & $0.01 \pm 0.002$ & $-0.23 \pm 0.01$ & $2.27 \pm 0.04$ & $0.29 \pm 0.01$ & $0.14 \pm 0.01$ & $0.15 \pm 0.01$ \\
210812A(1) & $58.79 \pm 1.72$ & $210.92 \pm 9.90$ & $0.17 \pm 0.02$ & $-5.04 \pm 0.02$ & $1.00 \pm 0.31$ & $2.05 \pm 0.14$ & $0.94 \pm 0.06$ & $1.11 \pm 0.08$ \\
210812A(2) & $5.79 \pm 0.08$ & $2779.11 \pm 267.84$ & $0.14 \pm 0.004$ & $15.32 \pm 0.004$ & $35.04 \pm 0.99$ & $3.32 \pm 0.11$ & $1.59 \pm 0.05$ & $1.73 \pm 0.06$ \\
\bottomrule
\end{longtable}

\subsection{Spectral Characteristics}
\subsubsection{spectral component analysis}
To examine the presence of thermal components in the spectra, we take GRB 200716C as an illustrative example, as shown in Figure~\ref{fig3:$DIC$_6grbs}. The first pulse is divided into six time bins and the second pulse into five time bins using the BBlocks (Bayesian Blocks) method. The number of Bayesian block segments obtained for each pulse is primarily determined by photon statistics and the chosen false-positive rate and S/N threshold. In practice, brighter pulses produce more Bayesian blocks; conversely, lower-count pulses yield fewer blocks that meet our spectral S/N criterion (S $>$ 5). This is to ensure that there are enough counts per bin to achieve a robust spectral fit. Consequently, it is expected that two pulses of the same burst may yield different numbers of valid time bins for spectral fitting.
Table~\ref{table1} summarizes the spectral fitting results together with the corresponding goodness-of-fit metrics. We begin by fitting the spectrum using the Band and CPL models. As shown in Figure~\ref{fig3:$DIC$_6grbs} and Table \ref{table1}, for the second time-resolved spectrum of the first pulse (0.1s to 0.27s), the DIC for the Band model is 803.16, whereas the DIC for the CPL model is 805.39. $\Delta DIC =DIC_{Band} - {DIC}_{CPL} = -2.23$, indicating positive evidence in favor of the Band model as the best fit for the second time-resolved spectrum of the first pulse. Subsequently, by incorporating a thermal component, we fit the spectra using the Band+BB and CPL+BB models. The DIC values are 749.02 for Band+BB and 743.48 for CPL+BB, yielding $\Delta$DIC = $DIC_{Band+BB}$ -${DIC}_{CPL+BB}$=$ 5.54$, which provides positive evidence in favor of the CPL+BB model. Finally, by comparing the DIC values of the two best-fitting models, we assess the presence of a thermal component. The DIC difference, $\Delta DIC_{best} = DIC_{best2} - DIC_{best1} = 59.68$, where $DIC_{best2}$ and $DIC_{best1}$ represent the DIC values of the best-fit models without and with the thermal component, respectively, provides strong evidence favoring the CPL+BB model. This suggests a prominent thermal contribution in the spectrum of this time bin. As shown in Figure~\ref{fig3:$DIC$_6grbs} for GRB 200716C, the $\Delta$DIC values for all time bins in both pulses exceed 10, indicating strong evidence for the presence of thermal components in each pulse. Applying the same analysis to the remaining five GRBs, the results also displayed in Figure~\ref{fig3:$DIC$_6grbs} demonstrate that the two pulses in each burst show similar time-resolved spectral behavior, with thermal components consistently present in both.

\begin{longtable}{cccccccc}
\caption{Spectral Fitting Result of GRB 200716C the first pulse}
\label{table1}\\
\hline
$t_{\text{start}}-t_{\text{end}}$ & $S$ & Model & $\alpha$ & $\beta$ & $E_p$ & kT & DIC\\
 s& & & & & keV & keV & \\
(1)&(2)&(3)&(4)&(5)&(6)&(7)&(8)\\
\hline
\endfirsthead
0.0--0.1 & 6.0 & band & $-0.80^{+0.27}_{-0.29}$ & $-2.09^{+0.30}_{-0.47}$ & $223.18^{+189.65}_{-62.40}$ & ... & 218.36 \\
 & 6.0 & cpl & $-0.82^{+0.27}_{-0.23}$ & ... & $252.41^{+148.52}_{-61.81}$ & ... & 220.03 \\
 & 6.0 & bandbb & $-0.94^{+0.41}_{-0.15}$ & $-2.23^{+0.37}_{-0.34}$ & $174.52^{+128.70}_{-70.53}$ & $91.83^{+1.32}_{-47.21}$ & 204.56 \\
 & 6.0 & cplbb & $-0.83^{+0.26}_{-0.27}$ & ... & $171.37^{+165.56}_{-47.88}$ & $91.83^{+2.65}_{-52.49}$ & 204.55 \\
0.1--0.27 & 21.0 & band & $-0.74^{+0.09}_{-0.11}$ & $-2.01^{+0.13}_{-0.40}$ & $596.72^{+213.88}_{-108.13}$ & ... & 803.16 \\
 & 21.0 & cpl & $-0.76^{+0.07}_{-0.11}$ & ... & $678.29^{+275.34}_{-103.29}$ & ... & 805.39 \\
 & 21.0 & bandbb & $-0.46^{+0.03}_{-0.39}$ & $-1.73^{+0.02}_{-0.56}$ & $161.33^{+552.58}_{-9.53}$ & $145.21^{+6.66}_{-96.89}$ & 749.02 \\
 & 21.0 & cplbb & $-0.49^{+0.07}_{-0.41}$ & ... & $169.14^{+718.56}_{-7.71}$ & $164.18^{+0.48}_{-121.81}$ & 743.48 \\
0.27--0.36 & 38.0 & band & $-0.49^{+0.05}_{-0.06}$ & $-2.70^{+0.22}_{-0.33}$ & $912.96^{+134.42}_{-87.37}$ & ... & 497.06 \\
 & 38.0 & cpl & $-0.52^{+0.05}_{-0.06}$ & ... & $1001.63^{+127.34}_{-80.76}$ & ... & 491.78 \\
 & 38.0 & bandbb & $-0.44^{+0.01}_{-0.17}$ & $-2.35^{+0.03}_{-0.63}$ & $529.10^{+566.73}_{-123.63}$ & $280.08^{+65.32}_{-219.79}$ & 479.74 \\
 & 38.0 & cplbb & $-0.53^{+0.04}_{-0.10}$ & ... & $1026.41^{+186.35}_{-78.10}$ & $110.48^{+27.95}_{-53.67}$ & 478.81 \\
0.36--0.47 & 36.0 & band & $-0.42^{+0.06}_{-0.14}$ & $-2.36^{+0.13}_{-0.40}$ & $302.82^{+56.12}_{-24.33}$ & ... & 596.63 \\
 & 36.0 & cpl & $-0.55^{+0.07}_{-0.07}$ & ... & $369.60^{+39.02}_{-33.08}$ & ... & 595.81 \\
 & 36.0 & bandbb & $-0.38^{+0.08}_{-0.17}$ & $-2.42^{+0.20}_{-0.32}$ & $247.35^{+90.61}_{-33.02}$ & $159.23^{+5.52}_{-112.79}$ & 577.64 \\
 & 36.0 & cplbb & $-0.31^{+0.03}_{-0.28}$ & ... & $217.52^{+166.04}_{-15.54}$ & $192.09^{+5.87}_{-144.20}$ & 562.95 \\
0.47--0.55 & 16.0 & band & $-0.60^{+0.19}_{-0.19}$ & $-2.49^{+0.18}_{-0.42}$ & $115.76^{+18.85}_{-13.74}$ & ... & 137.35 \\
 & 16.0 & cpl & $-0.67^{+0.21}_{-0.15}$ & ... & $128.25^{+16.74}_{-15.77}$ & ... & 134.58 \\
 & 16.0 & bandbb & $-0.69^{+0.29}_{-0.11}$ & $-2.56^{+0.22}_{-0.35}$ & $121.06^{+10.61}_{-24.32}$ & $54.87^{+22.32}_{-23.50}$ & 124.04 \\
 & 16.0 & cplbb & $-0.66^{+0.22}_{-0.14}$ & ... & $124.18^{+13.83}_{-19.14}$ & $56.66^{+29.88}_{-24.66}$ & 121.03 \\
0.55--1.0 & 15.0 & band & $-0.90^{+0.22}_{-0.18}$ & $-2.22^{+0.27}_{-0.33}$ & $129.55^{+33.57}_{-29.01}$ & ... & 1713.99 \\
 & 15.0 & cpl & $-0.97^{+0.13}_{-0.18}$ & ... & $149.66^{+46.70}_{-19.92}$ & ... & 1719.52 \\
 & 15.0 & bandbb & $-0.94^{+0.26}_{-0.11}$ & $-2.22^{+0.24}_{-0.38}$ & $124.91^{+22.80}_{-36.52}$ & $49.59^{+19.57}_{-20.44}$ & 1701.84 \\
 & 15.0 & cplbb & $-0.80^{+0.09}_{-0.31}$ & ... & $95.50^{+75.95}_{-5.12}$ & $103.73^{+12.33}_{-71.39}$ & 1696.33 \\
\hline
\end{longtable}

\begin{figure}[htbp]
\centering
\includegraphics[width=0.3\textwidth]{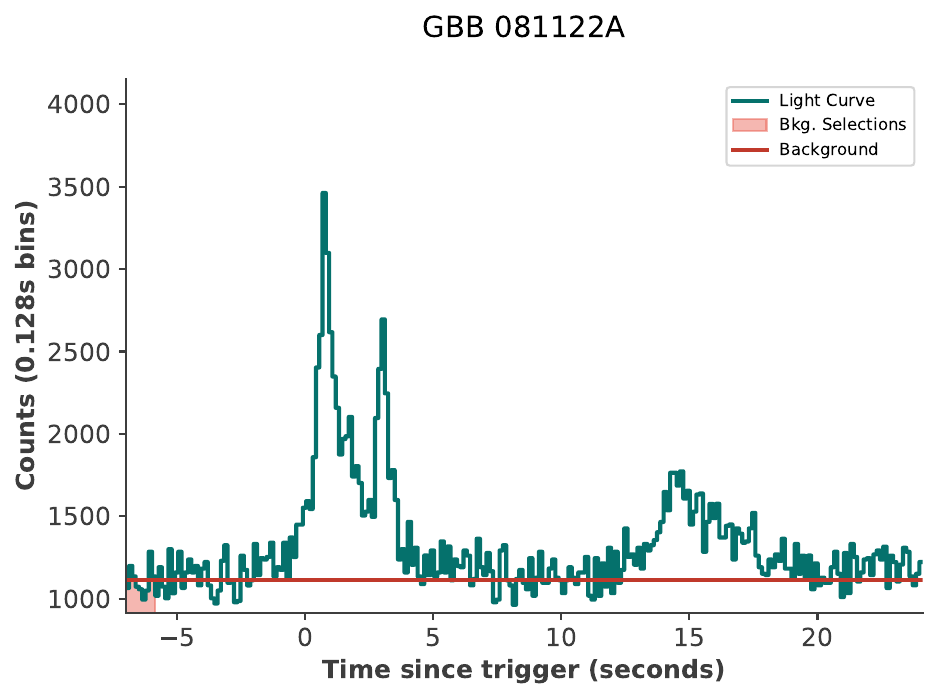}
\includegraphics[width=0.3\textwidth]{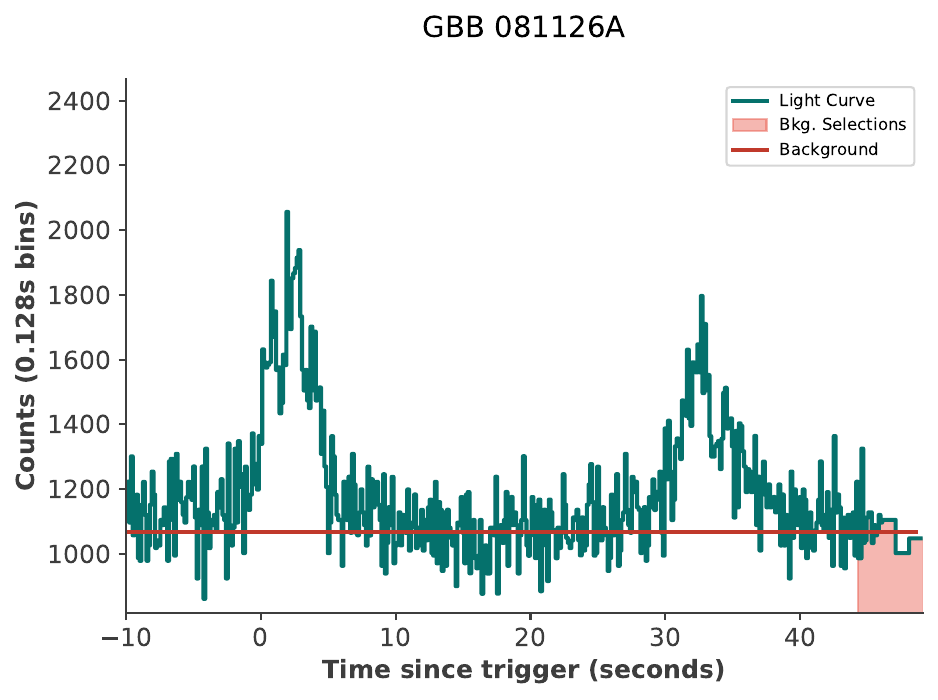}
\includegraphics[width=0.3\textwidth]{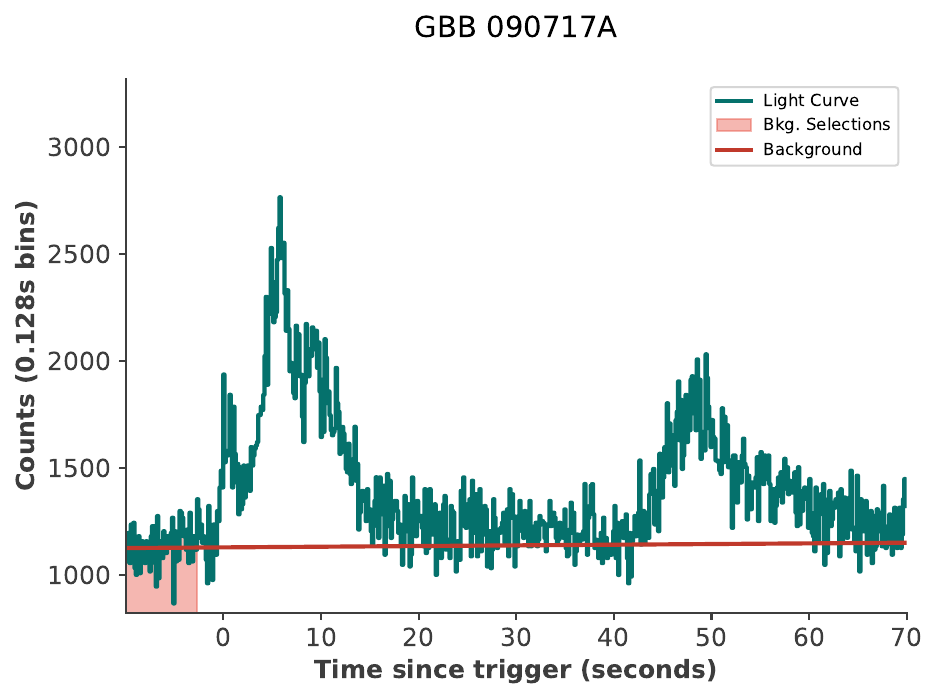}
\\
\includegraphics[width=0.3\textwidth]{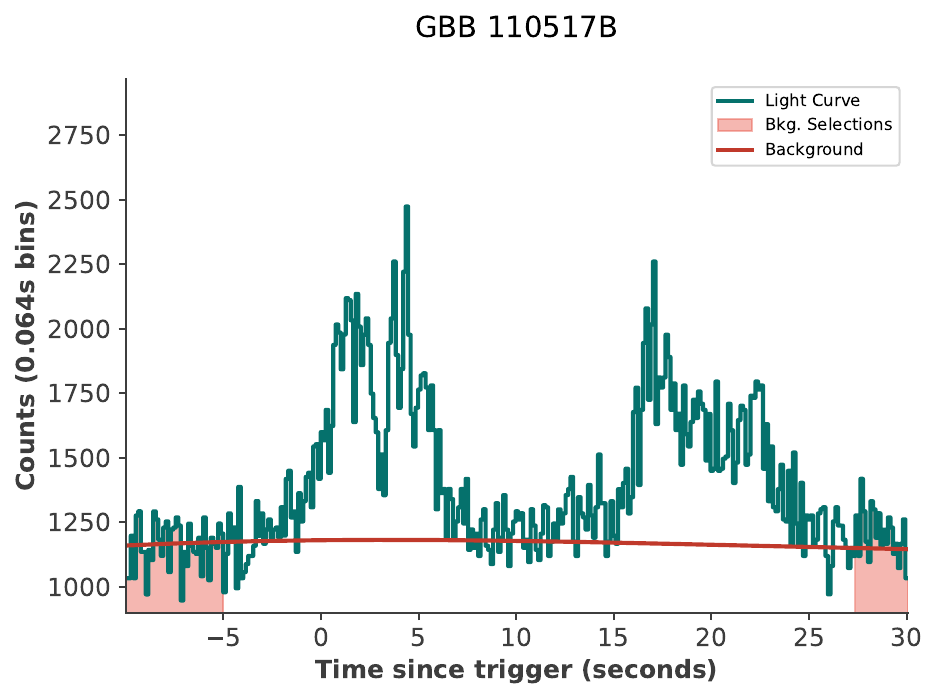}
\includegraphics[width=0.3\textwidth]{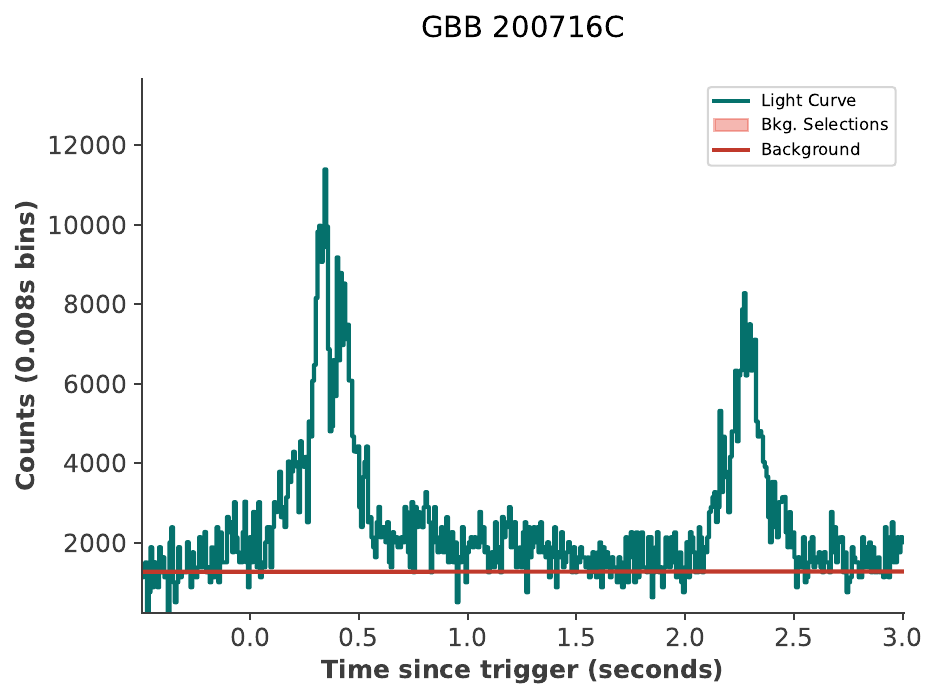}
\includegraphics[width=0.3\textwidth]{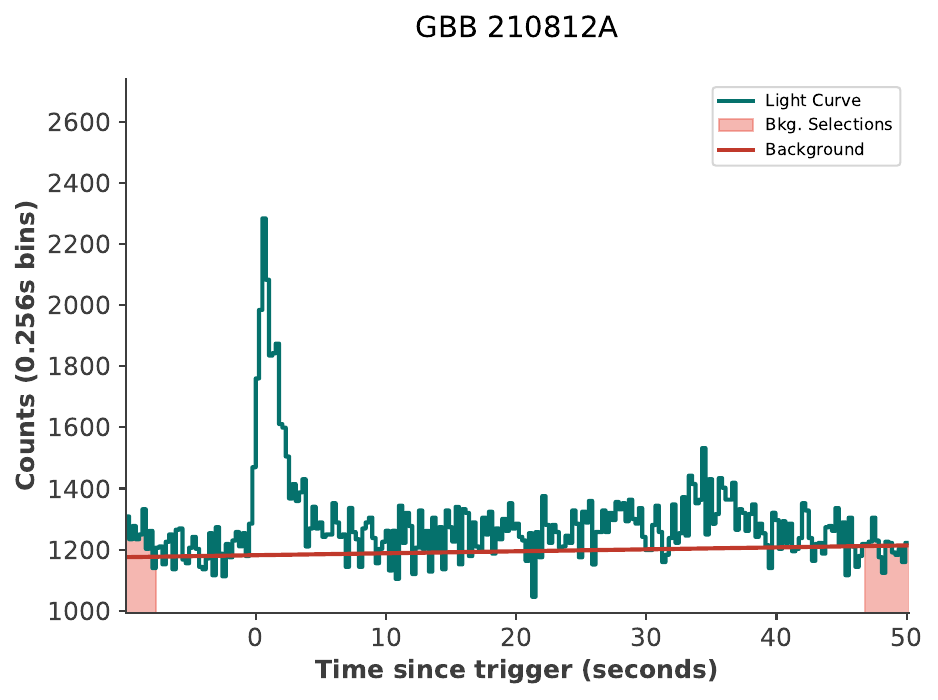}
\caption{Light curves of all GRBs in the corresponding time bins}
\label{fig2}
\end{figure}

\begin{figure}[htbp]
\centering
\includegraphics[width=0.3\textwidth]{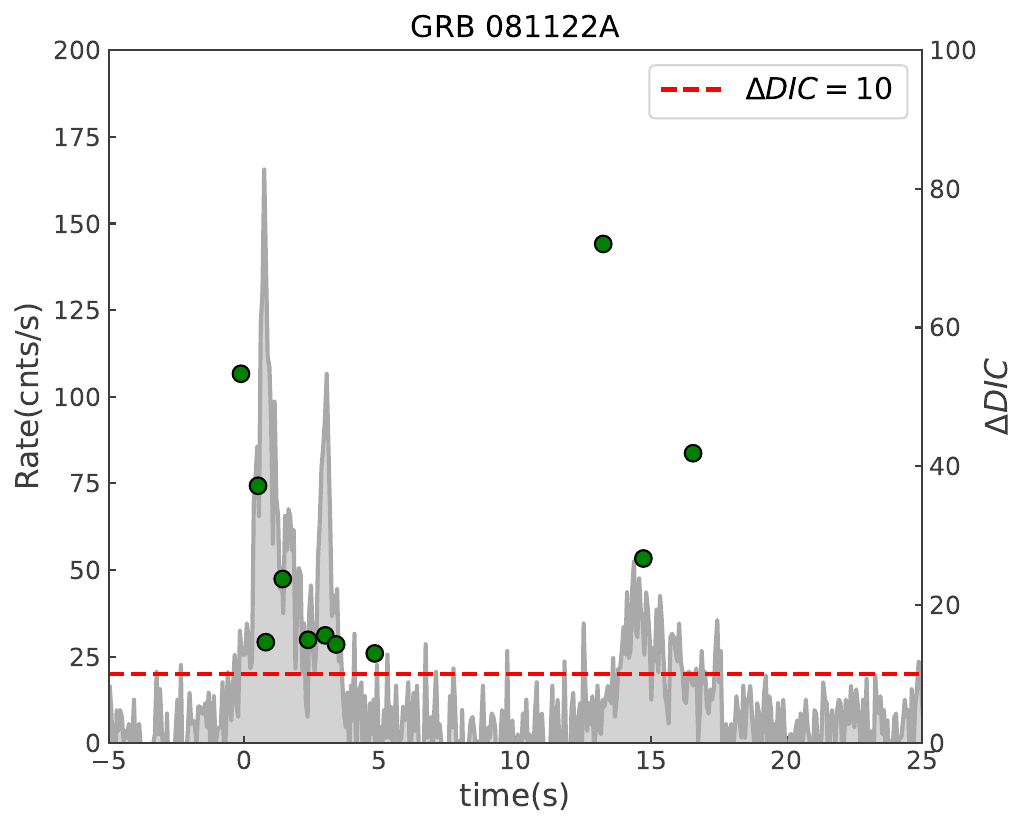}
\includegraphics[width=0.3\textwidth]{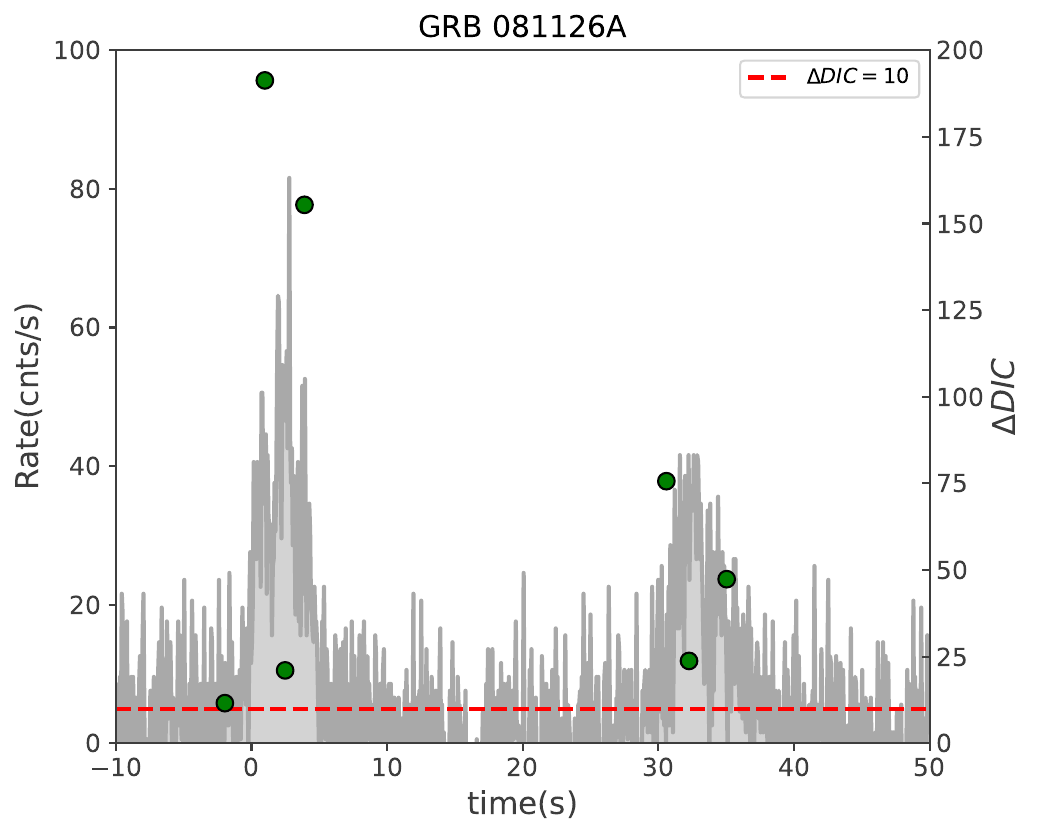}
\includegraphics[width=0.3\textwidth]{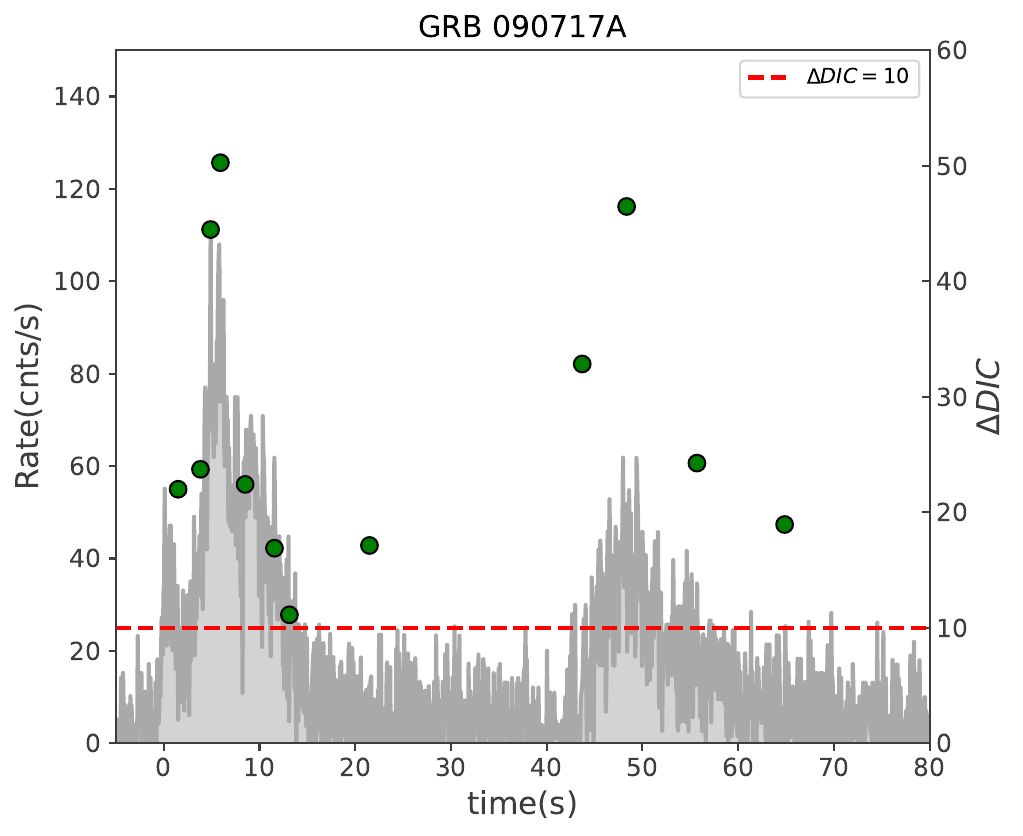}
\\
\includegraphics[width=0.3\textwidth]{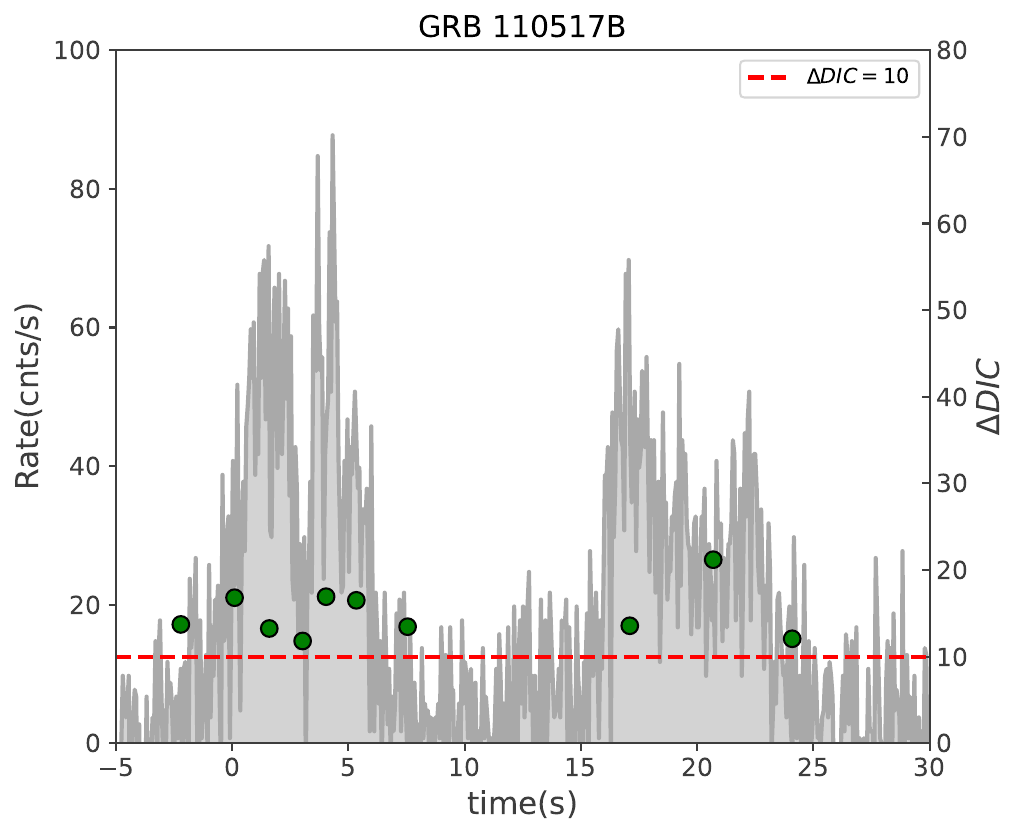}
\includegraphics[width=0.3\textwidth]{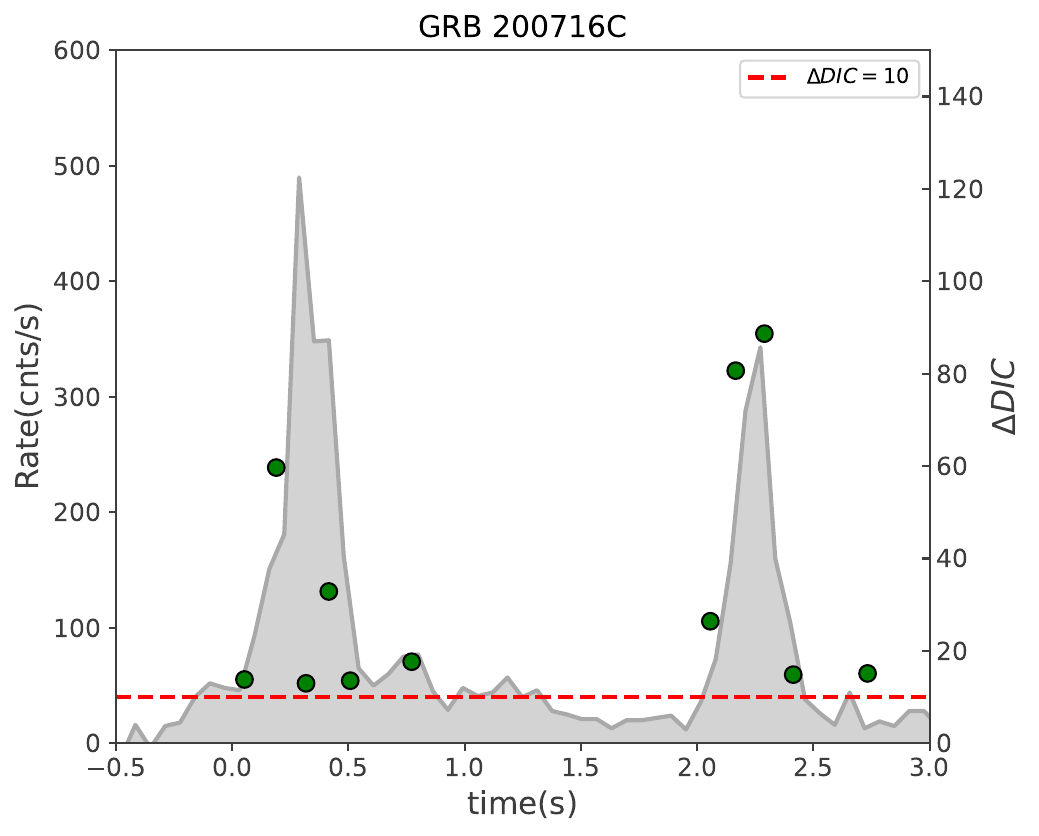}
\includegraphics[width=0.3\textwidth]{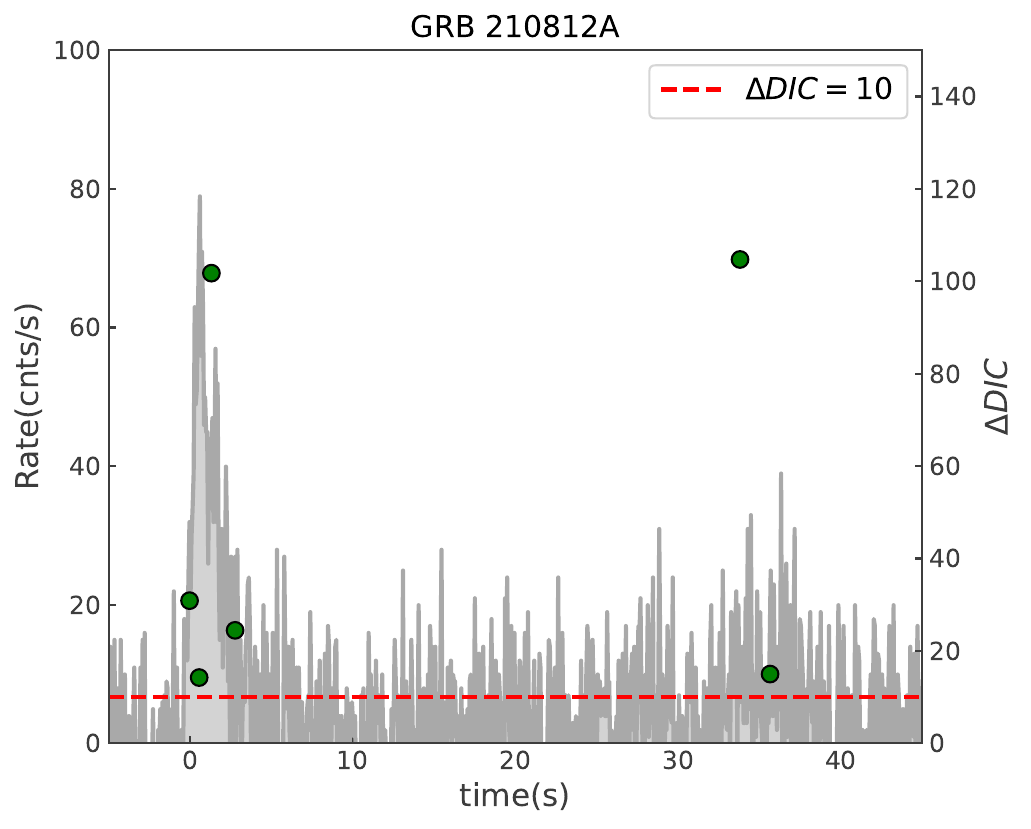}
\caption{The evolution of $\Delta DIC$ over time. The gray shading is the light curve, and the red dotted line indicates $\Delta DIC$ =10. Exceeding the red dashed line indicates compelling evidence of a thermal component}
\label{fig3:$DIC$_6grbs}
\end{figure}

\subsubsection{spectral parameters}
To further investigate the spectral characteristics of the pulses, we analyze the evolution of their spectral parameters. Previous studies have identified several patterns of spectral evolution: (1) the "hard-to-soft" (h.t.s) trend, (2) the "flux tracking" (f.t) trend, (3) the "soft-to-hard" (s.t.h) trend, and (4) other irregular or chaotic behaviors. When both the peak energy ($E_{p}$) and the low-energy spectral index ($\alpha$) follow the flux, the evolution is categorized as the "double tracking" mode \citep{2019ApJ...884..109L}. We conduct a detailed analysis of the spectral evolution for 12 pulses across 6 GRBs, with the results summarized in Table~\ref{tab4:grb_evolution}.

\begin{figure}[htbp]
\centering
\includegraphics[width=0.3\textwidth]{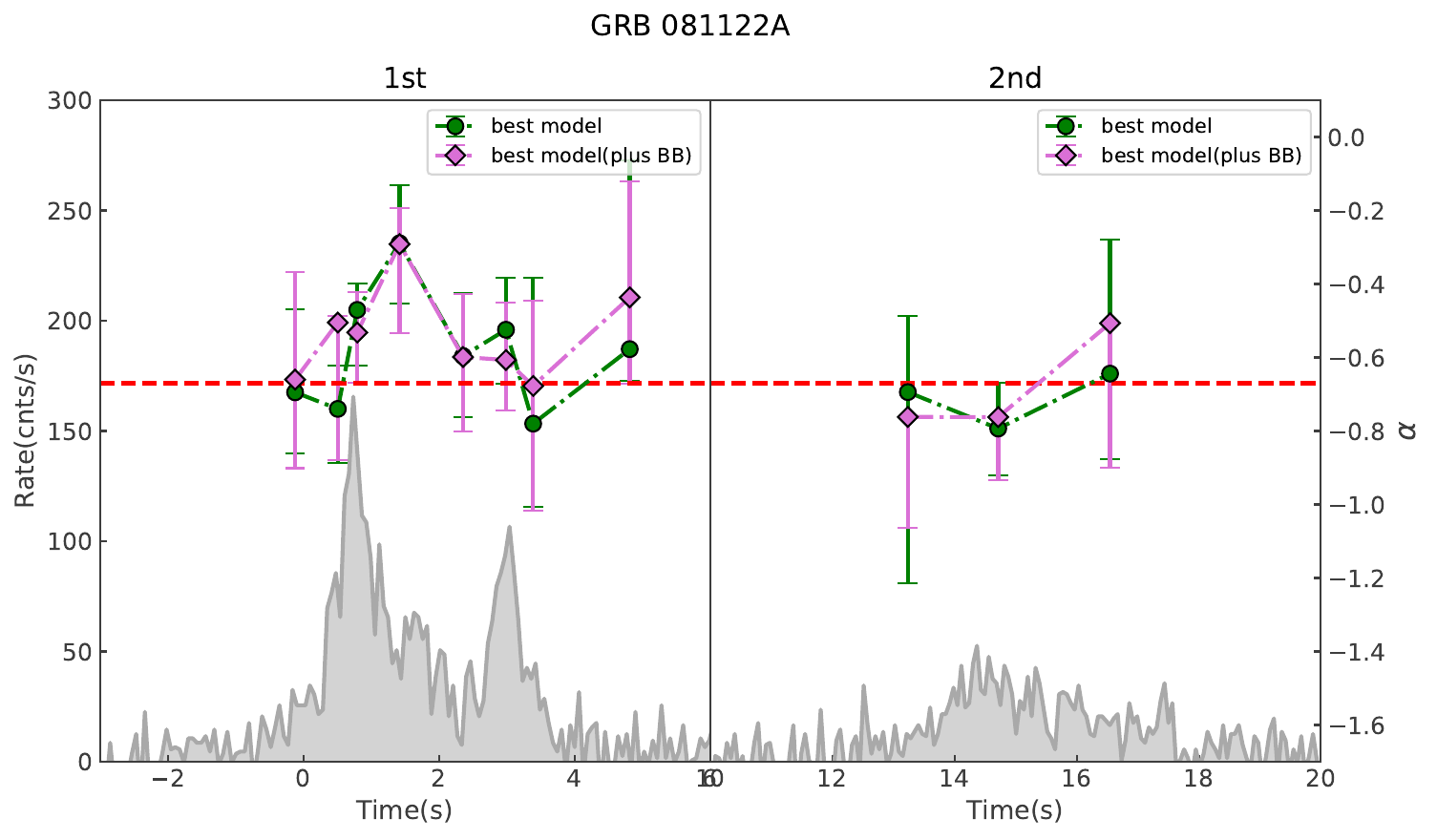}
\includegraphics[width=0.3\textwidth]{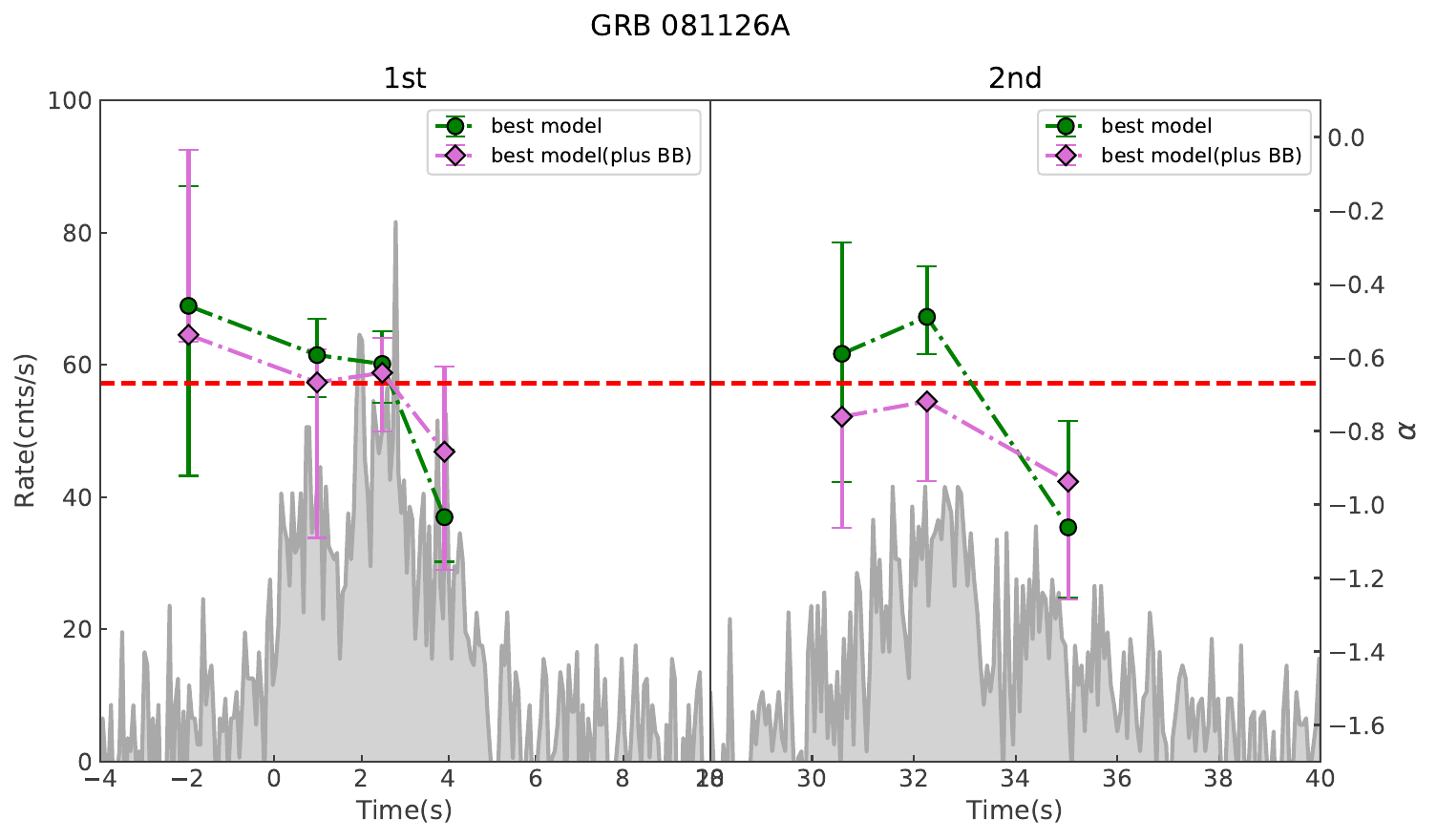}
\includegraphics[width=0.3\textwidth]{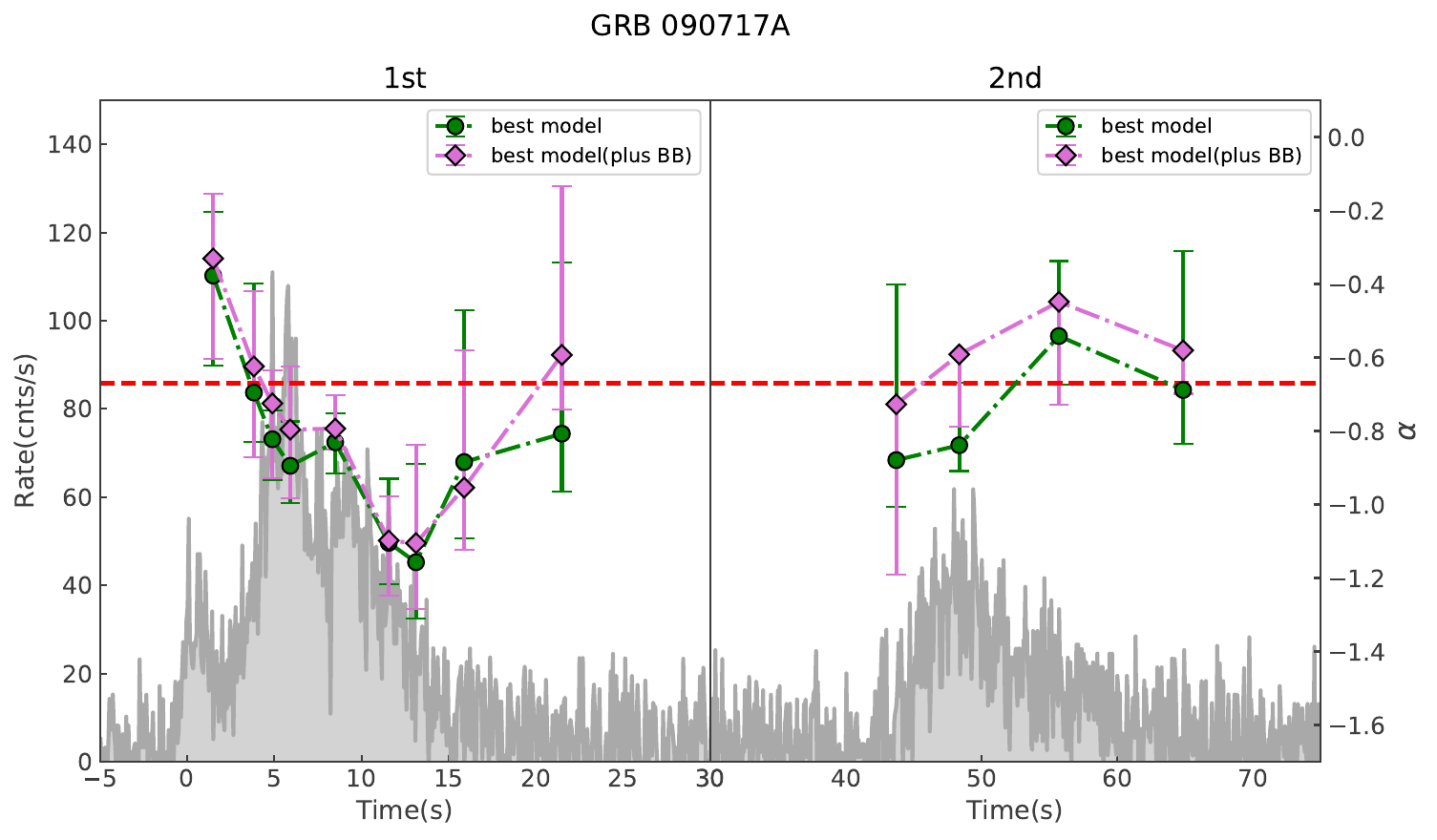}
\\
\includegraphics[width=0.3\textwidth]{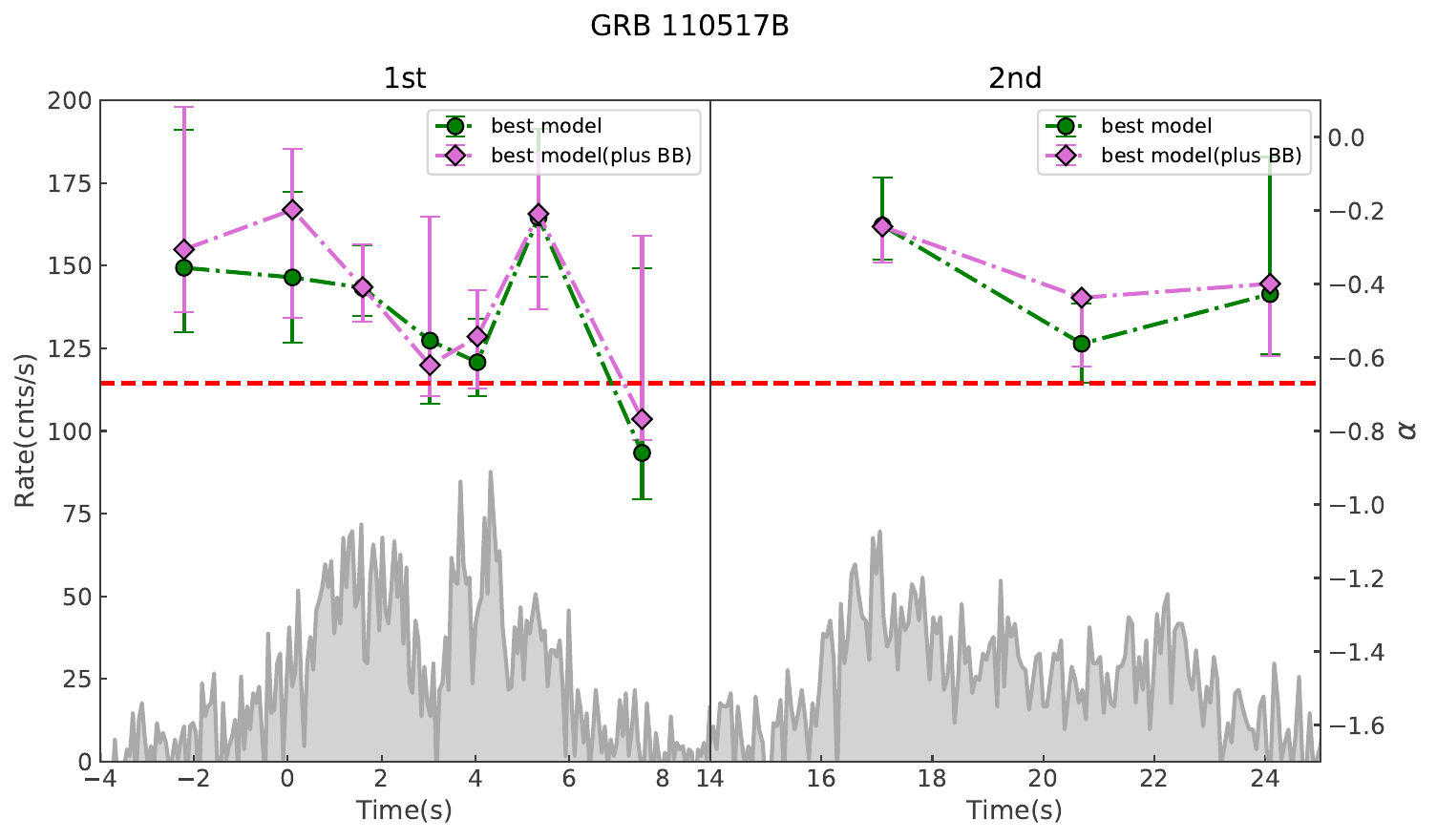}
\includegraphics[width=0.3\textwidth]{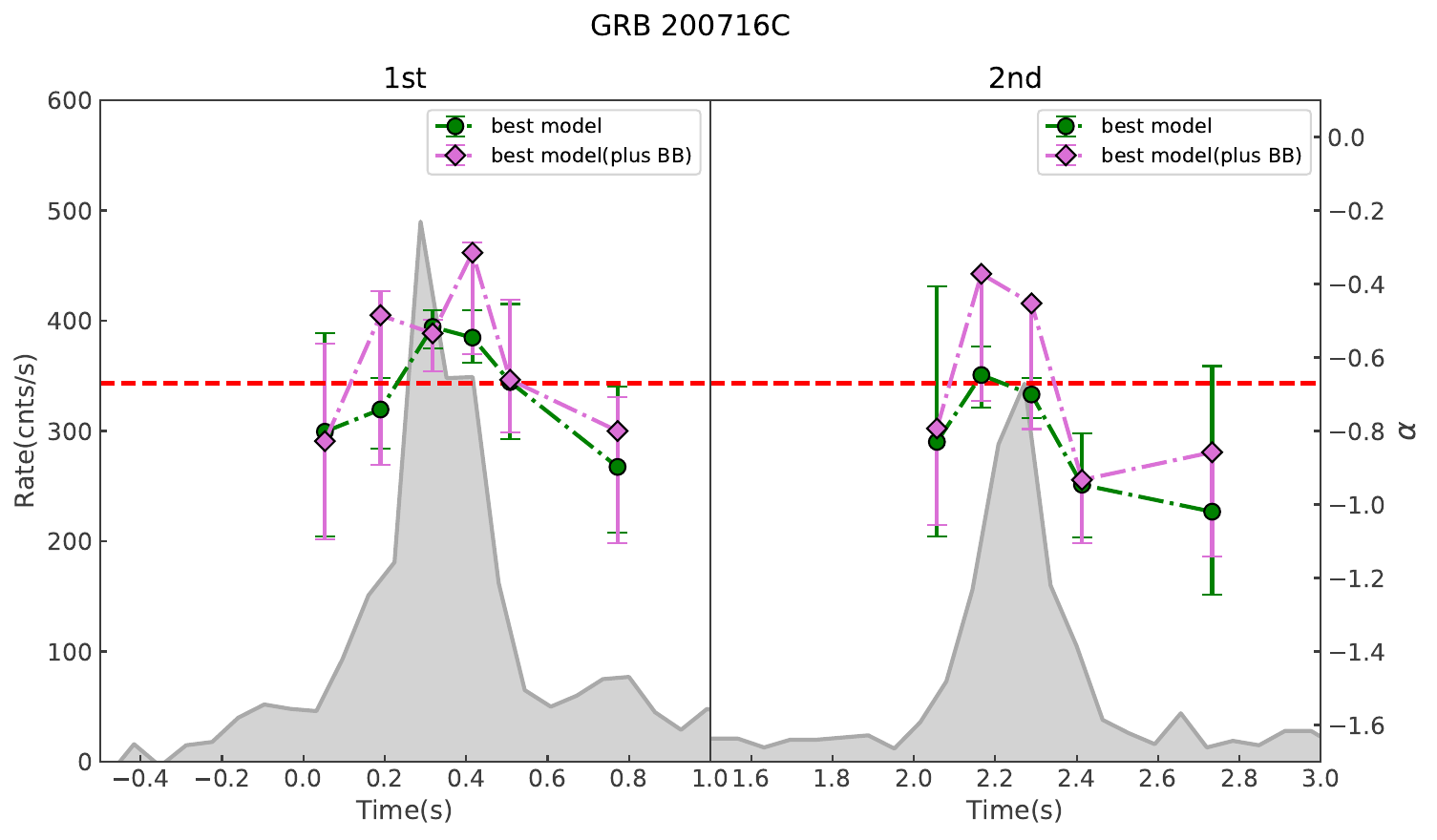}
\includegraphics[width=0.3\textwidth]{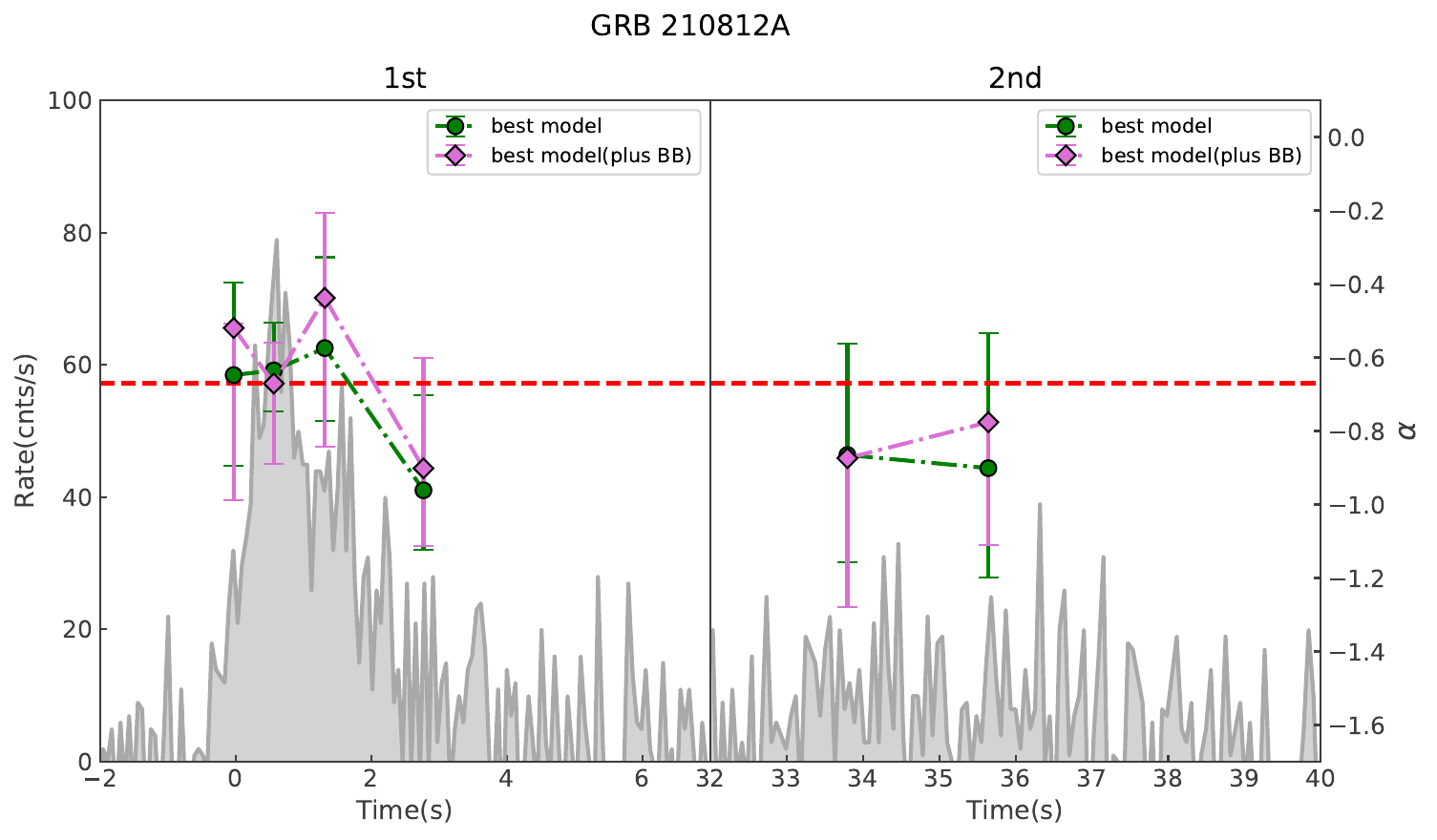}
\caption{The spectral parameter $\alpha$ fitted with the best model is shown as a function of time. The green and purple regions represent the best-fitting model without a BB component ("best model") and with a BB component ("best model + BB"), respectively. The red dashed line indicates $\alpha=-0.67$ }
\label{fig4:alpha_6grbs}
\end{figure}

The evolution of  $\alpha$  is shown in Figure \ref{fig4:alpha_6grbs}, while the evolution of $E_{p}$ is illustrated in Figure \ref{fig5:E_{p}}. The evolution of both  $\alpha$  and $E_{p}$ with time is summarized in Table \ref{tab4:grb_evolution}. Among the sample, all pulses—except for the second pulse of GRB 210812A—exhibit at least one time bin where the spectral parameters exceed the synchrotron death line ($\alpha = -2/3$; \citep{1998ApJ...506L..23P}). Four GRBs show consistent  $\alpha$ evolution patterns between the two pulses: GRB 081122A, GRB 081126A, GRB 110517B, and GRB 200716C. Specifically, GRB 081122A exhibits a s.t.h evolution pattern; GRB 081126A and GRB 110517B both display hard-to-soft evolution; and GRB 200716C presents a f.t evolution pattern. The evolution of $E_{p}$ in all pulses follows a f.t pattern. GRB 200716C is the only burst in the sample that shows a double-tracking spectral evolution in both $\alpha$ and $E_{p}$. These findings reveal that the spectral evolution of both $\alpha$ and $E_{p}$ in GRB 081122A, GRB 081126A, GRB 110517B, and GRB 200716C are highly consistent between the two pulses, suggesting internal similarity within each burst. 
 
Additionally, we compare the time-integrated spectral parameters $\alpha$ and $E_{p}$ of the two pulses within each burst, as shown in Table~\ref{tab4:grb_evolution}. Both the Band and CPL models are used for spectral fitting, and the best-fit model is selected based on the DIC. Taking GRB 110517B as an example, the spectral parameters of the two pulses are consistent within the 1$\sigma$ confidence level, as follows: $\alpha_1 = -0.58^{+0.09}_{-0.06}$, $E_{p1} = 115^{+6}_{-5}$ keV, $\alpha_2 = -0.52^{+0.10}_{-0.07}$, and $E_{p2} = 114^{+6}_{-5}$ keV. The same analysis is also conducted for the remaining bursts. The spectral parameters of all GRBs in both pulses are consistent within the 1$\sigma$ confidence level, except for GRB 200716C, whose $E_p$ parameter shows consistency within the 2$\sigma$ confidence level. For all GRBs, the spectral parameters of their two pulses are consistent within the 2$\sigma$ confidence level.

\begin{figure}[htbp]
\centering
\includegraphics[width=0.3\textwidth]{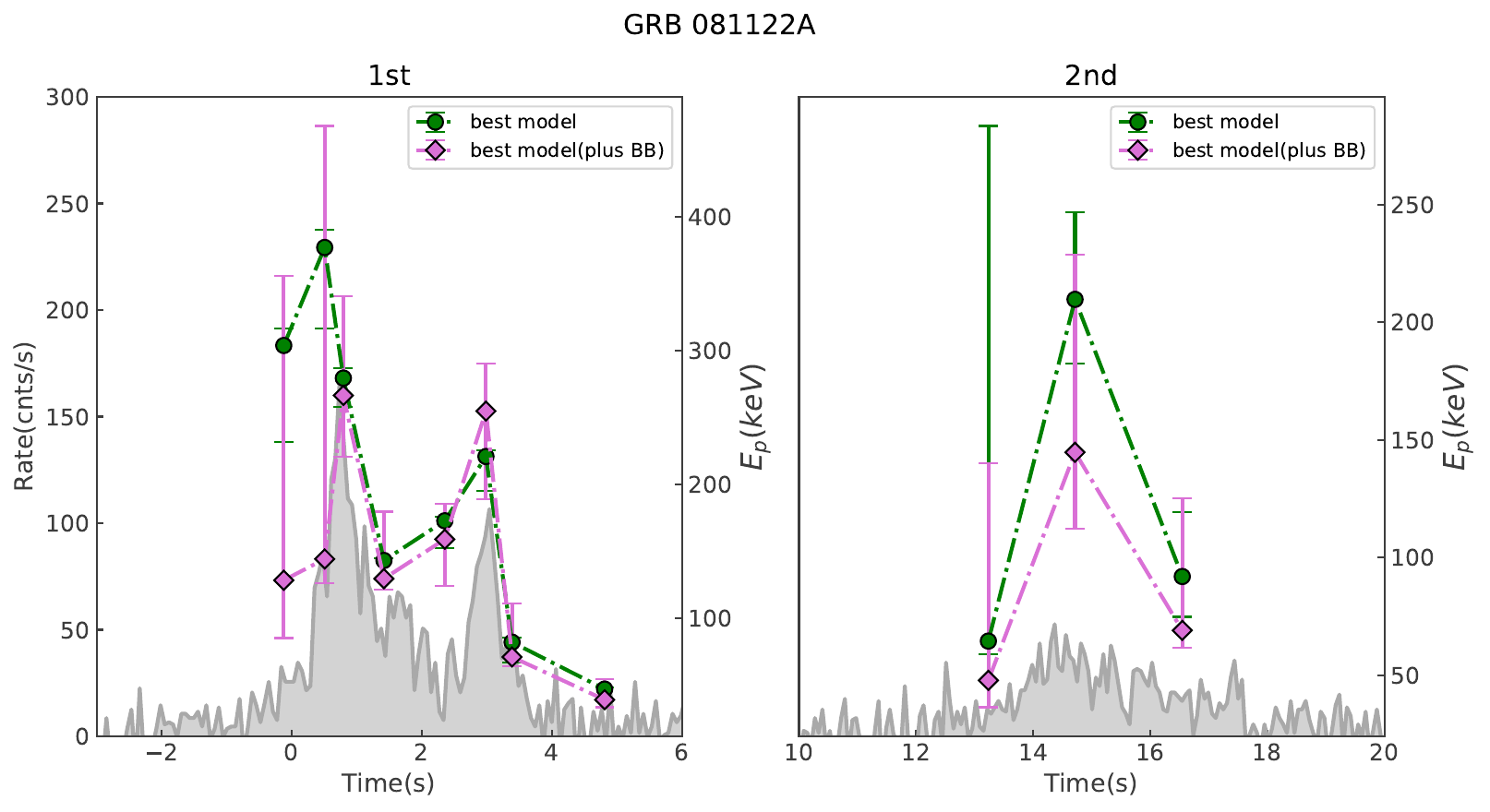}
\includegraphics[width=0.3\textwidth]{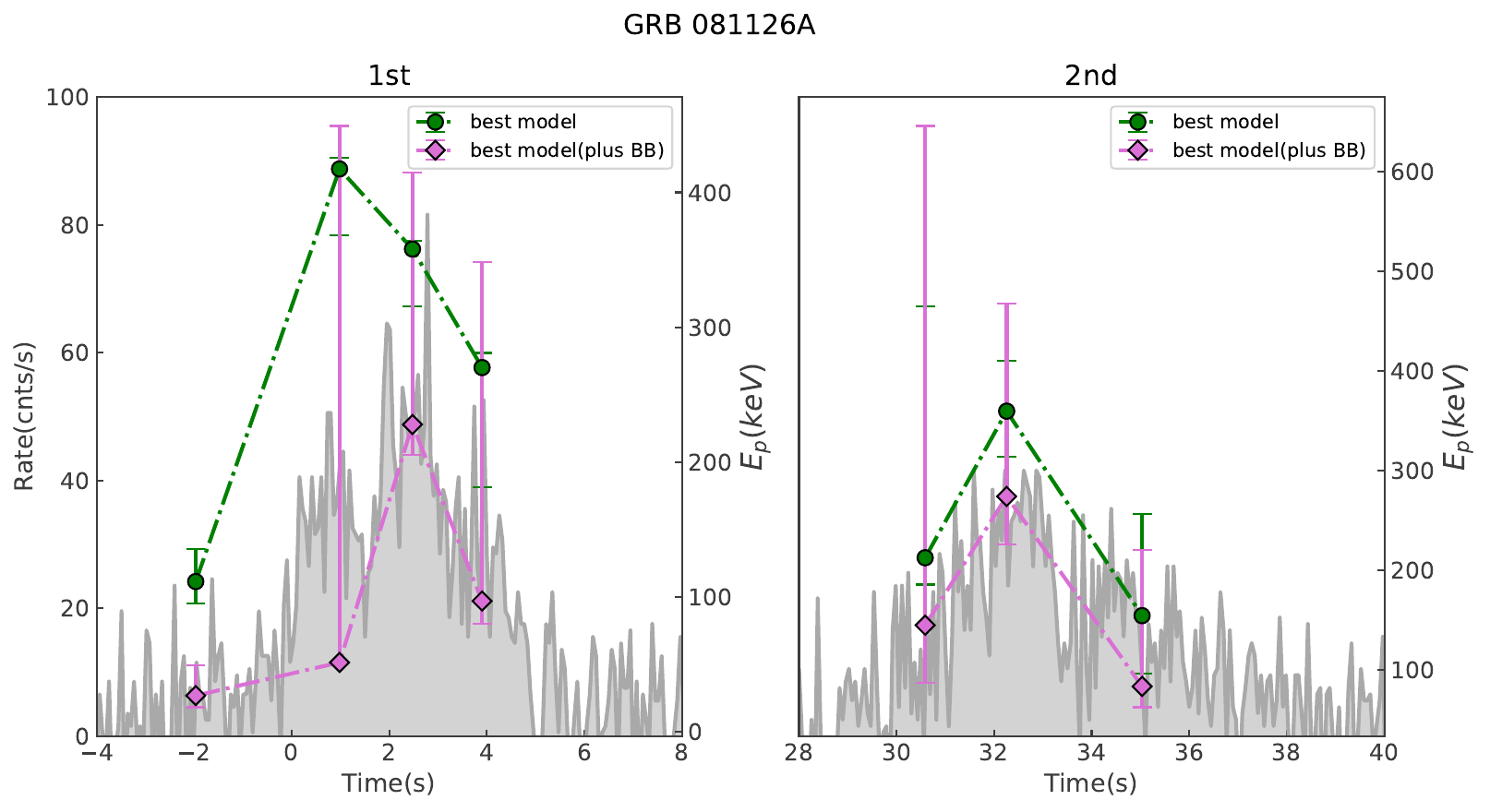}
\includegraphics[width=0.3\textwidth]{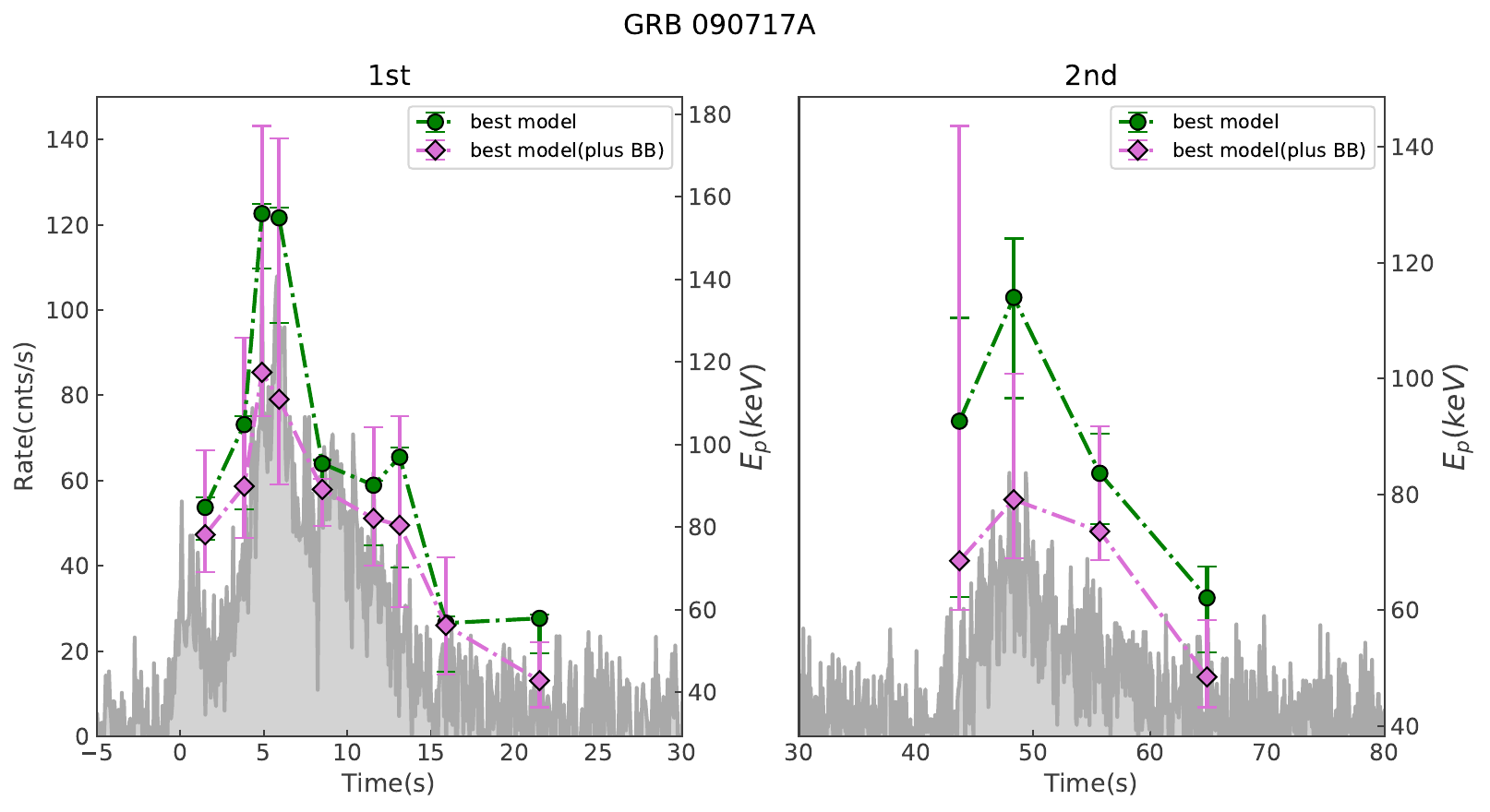}
\\
\includegraphics[width=0.3\textwidth]{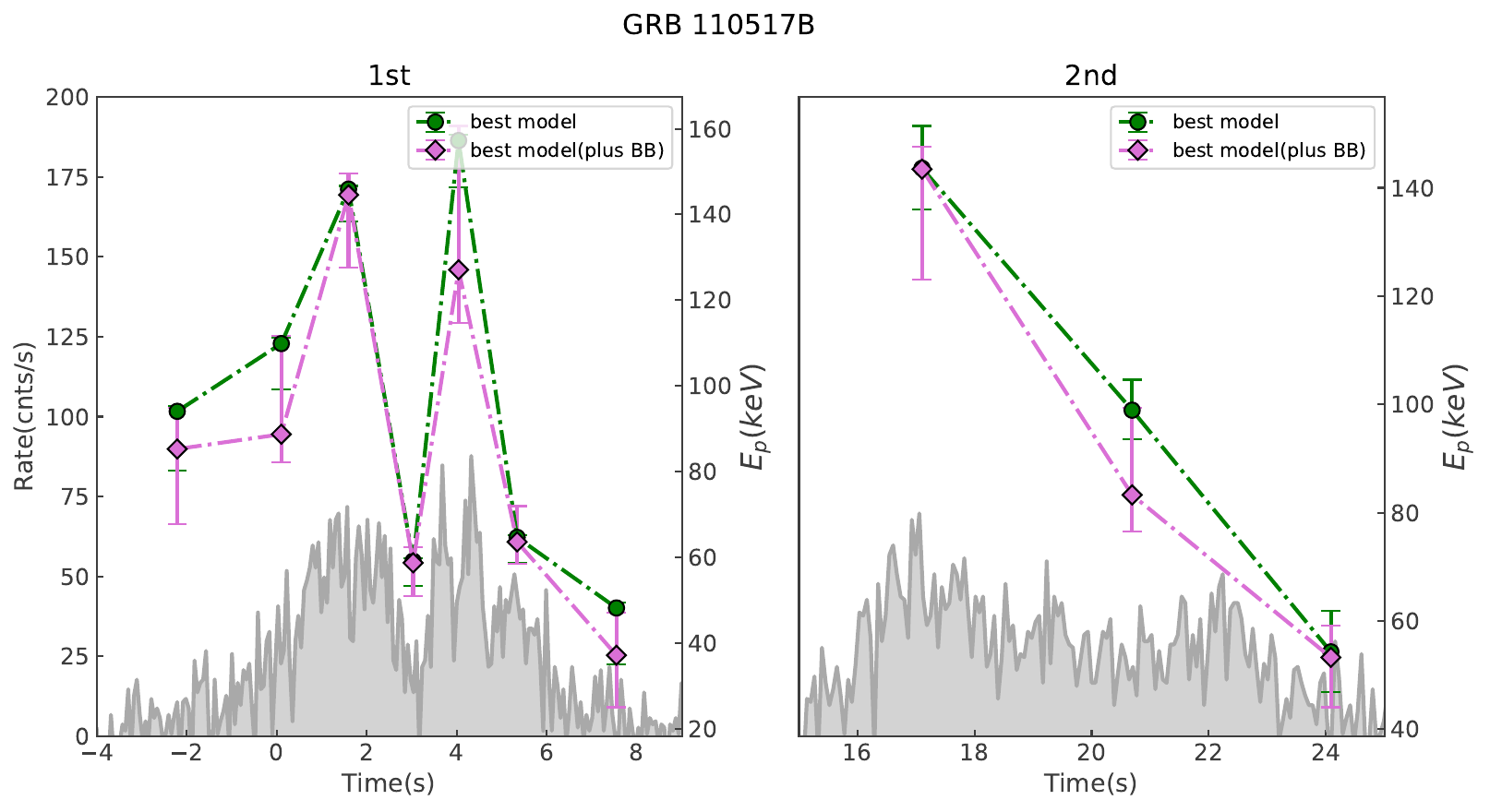}
\includegraphics[width=0.3\textwidth]{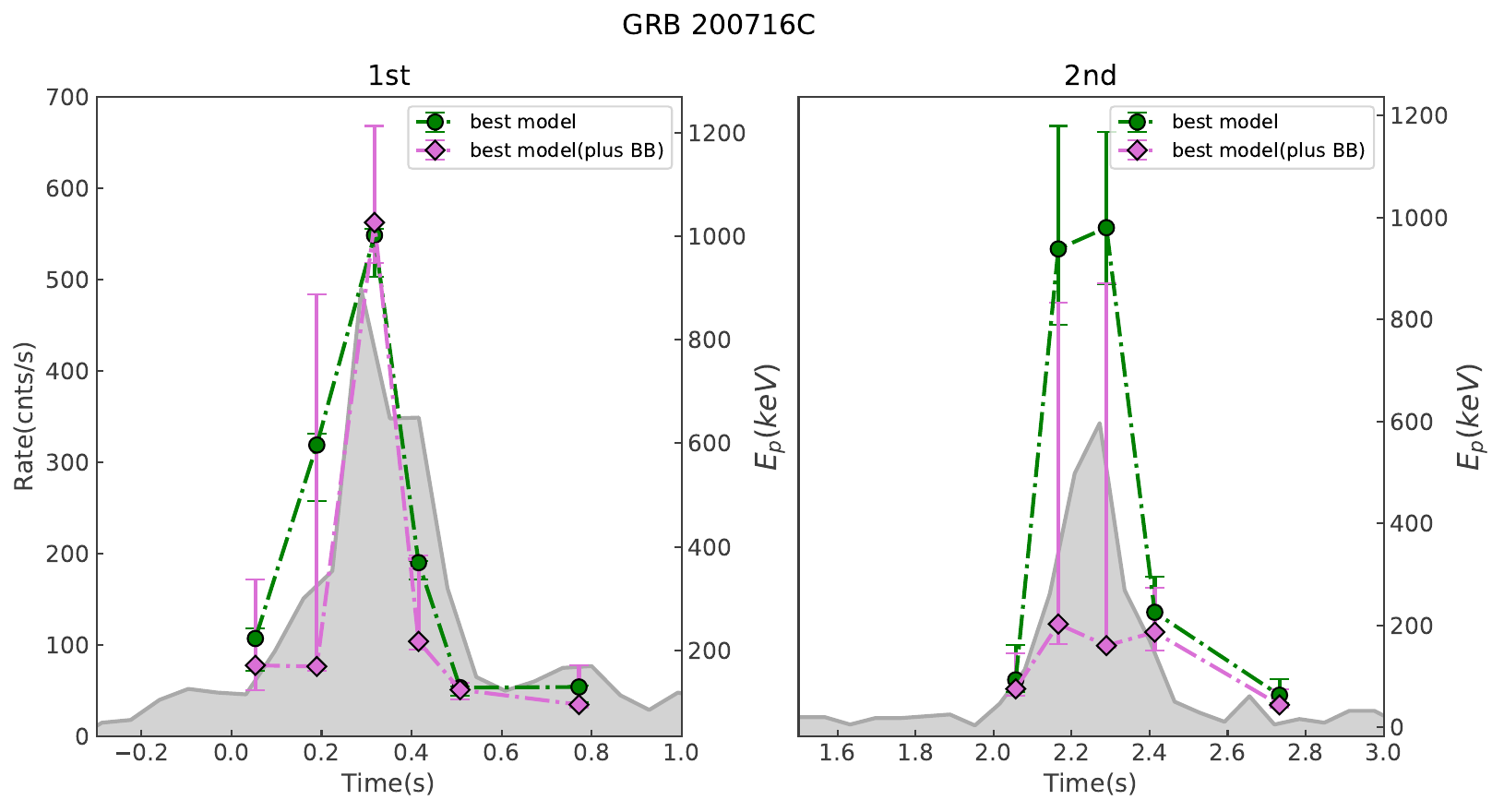}
\includegraphics[width=0.3\textwidth]{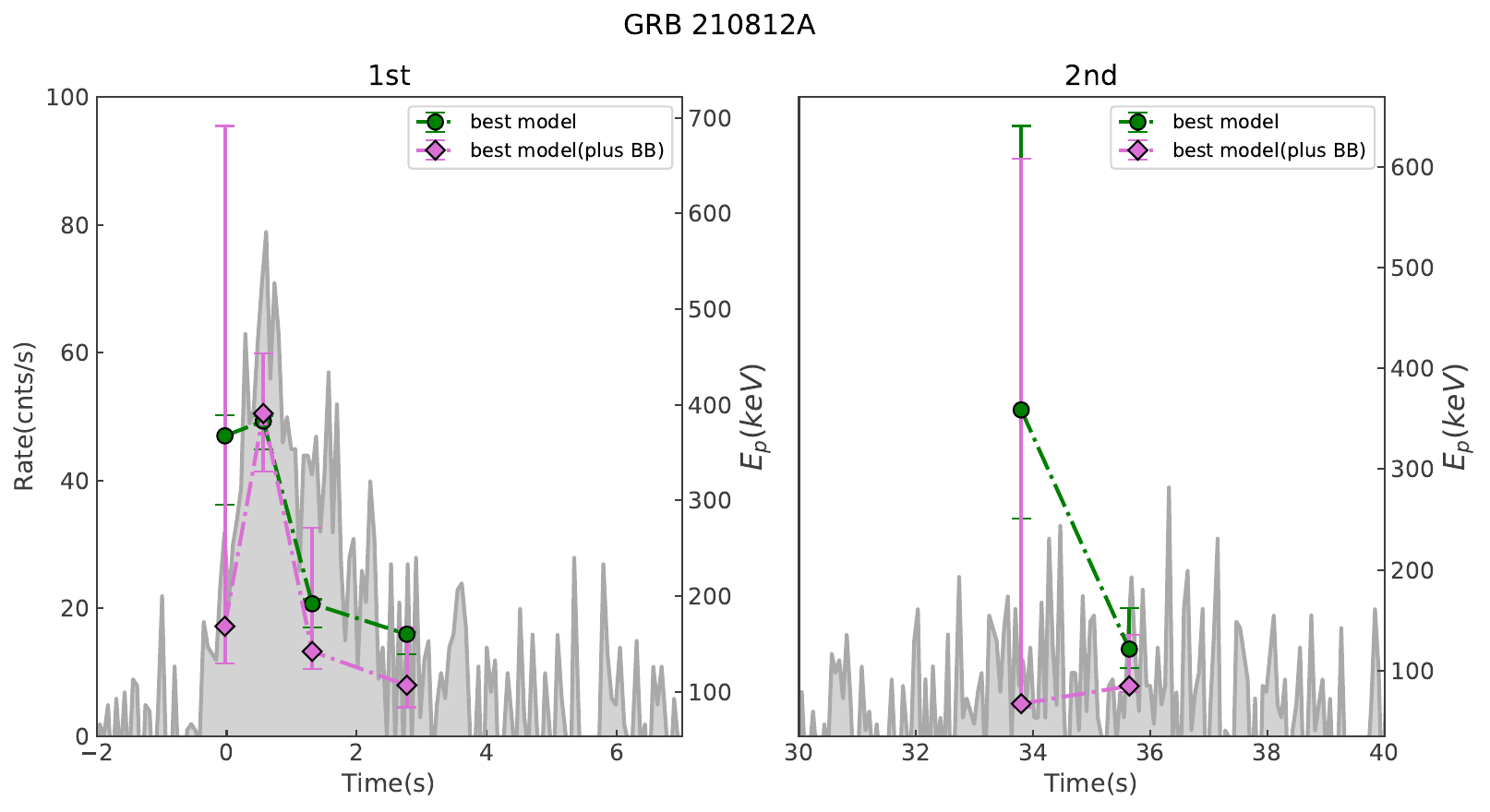}
\caption{$E_{p}$ evolution over time. The green part and the purple part represent the best model without BB component (best model) and the best model with BB component (best model+BB), respectively.}
\label{fig5:E_{p}}
\end{figure}

\begin{table}[htbp]
\centering
\caption{Spectral Parameters and evolution}
\label{tab4:grb_evolution}
\begin{tabular}{lccccccccc}
\toprule
\multirow{2}{*}{GRB Name} & \multicolumn{2}{c}{$\alpha$ Evolution} & \multicolumn{2}{c}{$E_p$ Evolution} & \multirow{2}{*}{Best Model} & \multicolumn{2}{c}{$\alpha$ Parameter} & \multicolumn{2}{c}{$E_p$ (keV)} \\
\cmidrule(lr){2-3} \cmidrule(lr){4-5} \cmidrule(lr){7-8} \cmidrule(lr){9-10}
 & 1st & 2nd & 1st & 2nd & & 1st & 2nd & 1st & 2nd \\
\midrule
GRB081122A  & s.t.h & s.t.h & f.t & f.t & Band & $-0.59^{+0.13}_{-0.10}$ & $-0.73 ^{+0.21}_{-0.29}$ & $180 ^{+32}_{-22}$ & $114 ^{+68}_{-20}$ \\
GRB081126A  & h.t.s & h.t.s & f.t & f.t & CPL  & $-0.89 ^{+0.08}_{-0.08}$ & $-0.86 ^{+0.10}_{-0.09}$ & $458 ^{+102}_{-66}$ & $319 ^{+71}_{-47}$ \\
GRB090717A  & h.t.s & s.t.h & f.t & f.t & Band & $-0.85^{+0.12}_{-0.08}$ & $-0.63 ^{+0.18}_{-0.11}$ & $98 ^{+13}_{-11}$  & $79 ^{+9}_{-8}$  \\
GRB110517B  & h.t.s & h.t.s & f.t & f.t & CPL  & $-0.58 ^{+0.09}_{-0.06}$ & $-0.52 ^{+0.10}_{-0.07}$ & $115 ^{+6}_{-5}$ & $114 ^{+6}_{-5}$ \\
GRB200716C  & f.t   & f.t   & f.t & f.t & CPL  & $-0.93 ^{+0.04}_{-0.03}$ & $-0.97 ^{+0.04}_{-0.05}$ & $499 ^{+61}_{-48}$ & $826 ^{+173}_{-126}$ \\
GRB210812A  & h.t.s. & --     & f.t &--  & CPL  & $-1^{+0.10}_{-0.09}$ & $-1.20 ^{+0.25}_{-0.04}$ & $294 ^{+39}_{-29}$ & $278 ^{+36}_{-85}$ \\
\bottomrule
\end{tabular}
\end{table}
\subsubsection{Distributions of spectral parameters}
\begin{figure}[htbp]
\centering
\includegraphics[width=\textwidth]{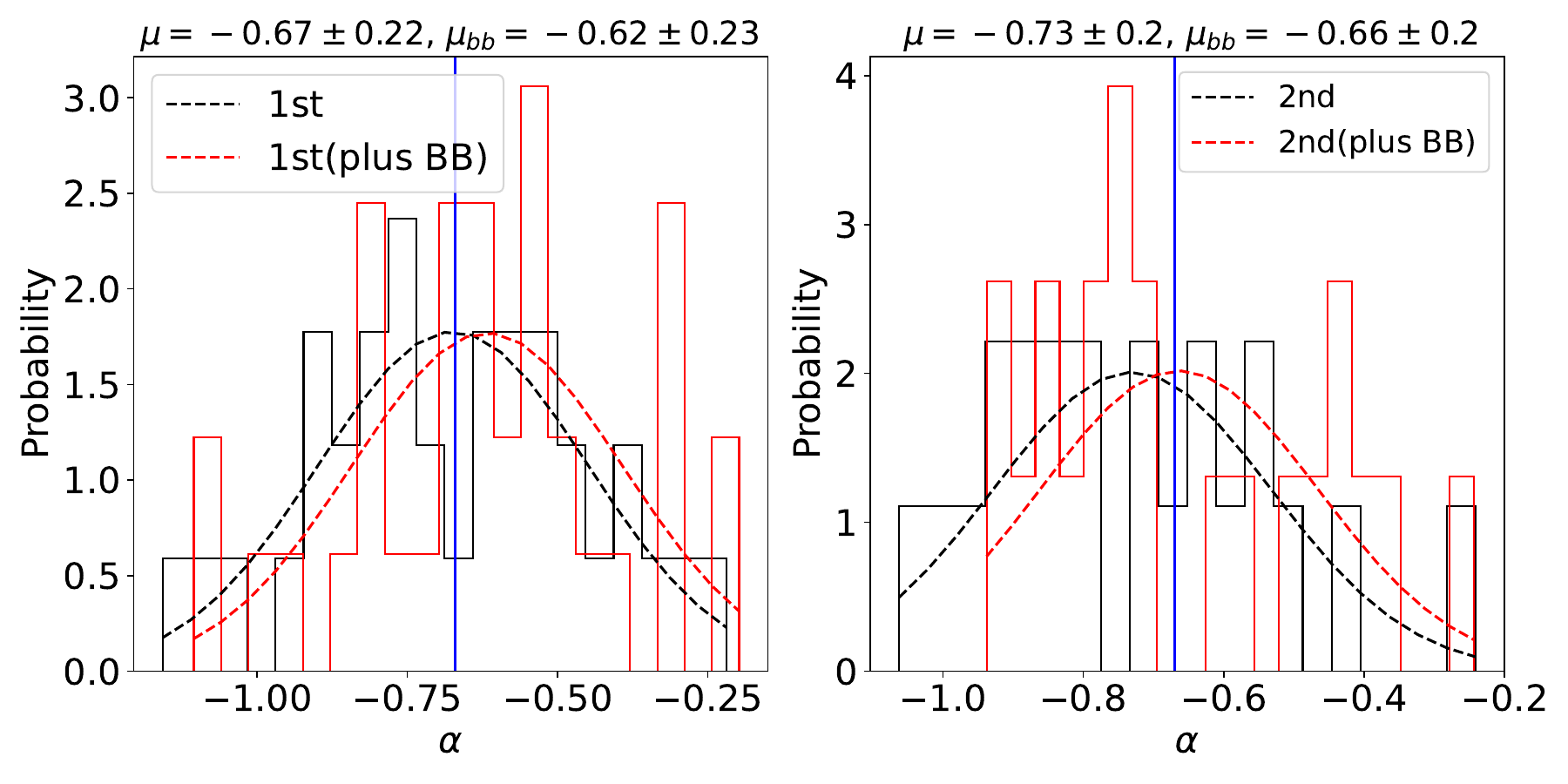}
\\
\vspace{0.3cm} 
\includegraphics[width=\textwidth]{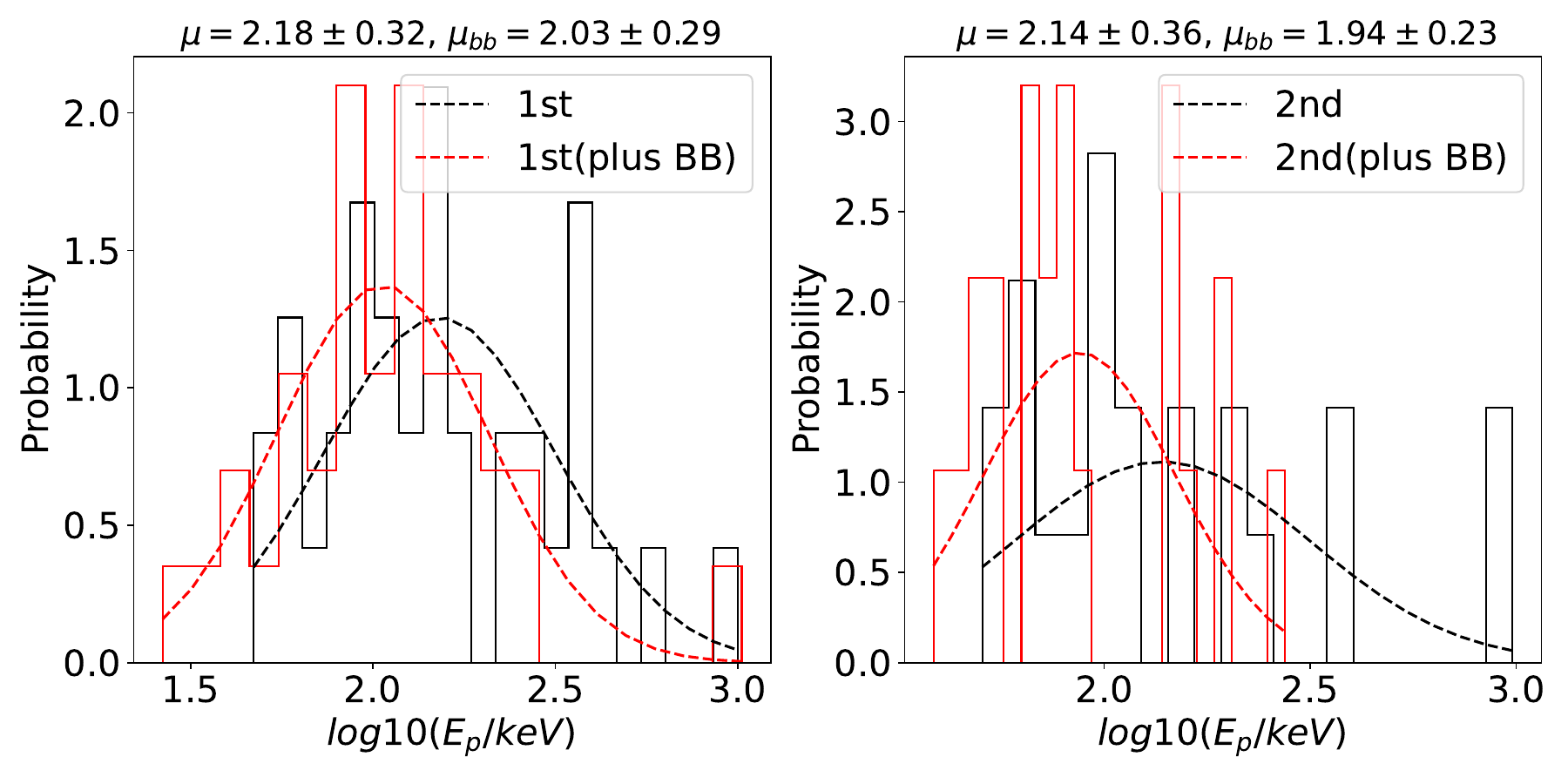}
\caption{The top panels show the distribution of $\alpha$, while the bottom panels show the distribution of $E_p$. The left and right panels correspond to the first and second pulses of the sample, respectively. The blue solid line represents $\alpha = -0.67$.}
\label{fig6}
\end{figure}

In the top panels of Figure ~\ref{fig6}, we present the distribution of $\alpha$ for the two pulses of each of the six GRBs in the sample. We apply Gaussian function fitting to obtain the mean and standard deviation of the spectral index $\alpha$. For the first pulses of all GRBs, the average $\alpha$ is $-0.67 \pm 0.22$, which shifts to $-0.62 \pm 0.23$ after including the BB component. For the second pulses, $\alpha$ is $-0.73 \pm 0.20$, and becomes $-0.66 \pm 0.20$ after the BB component is added. These statistical results indicate that the first pulses generally exhibit harder $\alpha$ values compared to the second pulses. Additionally, a larger fraction of time-resolved spectra in the first pulses exceed the "synchrotron death line" relative to the second pulses. In both cases, the inclusion of the BB component tends to harden the $\alpha$ values.

In the bottom panel of Figure \ref{fig6}, we present the distribution of $E_p$ in the overall sample. Similarly, we fit the $E_p$ distributions of all pulses in the sample using Gaussian functions. For the first pulses of all GRBs, we find $\log(E_p) = 2.18 \pm 0.32$, which shifts to $\log(E_p) = 2.03 \pm 0.29$ after incorporating the BB component. For the second pulses, $\log(E_p) = 2.14 \pm 0.32$, decreasing to $\log(E_p) = 1.94 \pm 0.32$ with the BB component. These results suggest that, for most time-resolved spectra, the $E_p$ values in the first pulses are generally lower than those in the second pulses. Additionally, the inclusion of the BB component leads to a slight reduction in $E_p$ values.
 
\begin{figure}[htbp]
\centering
\includegraphics[width=0.3\textwidth]{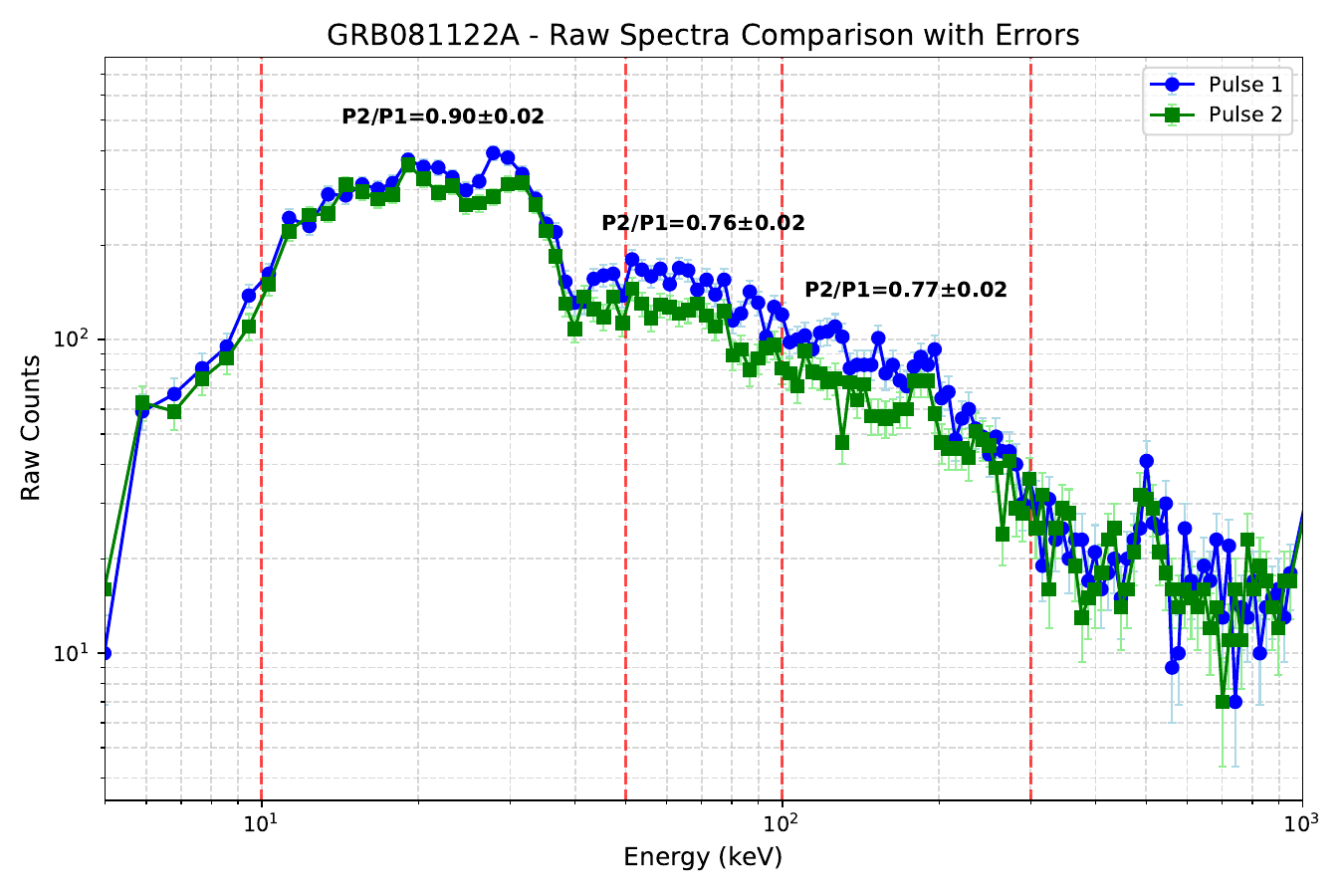}
\includegraphics[width=0.3\textwidth]{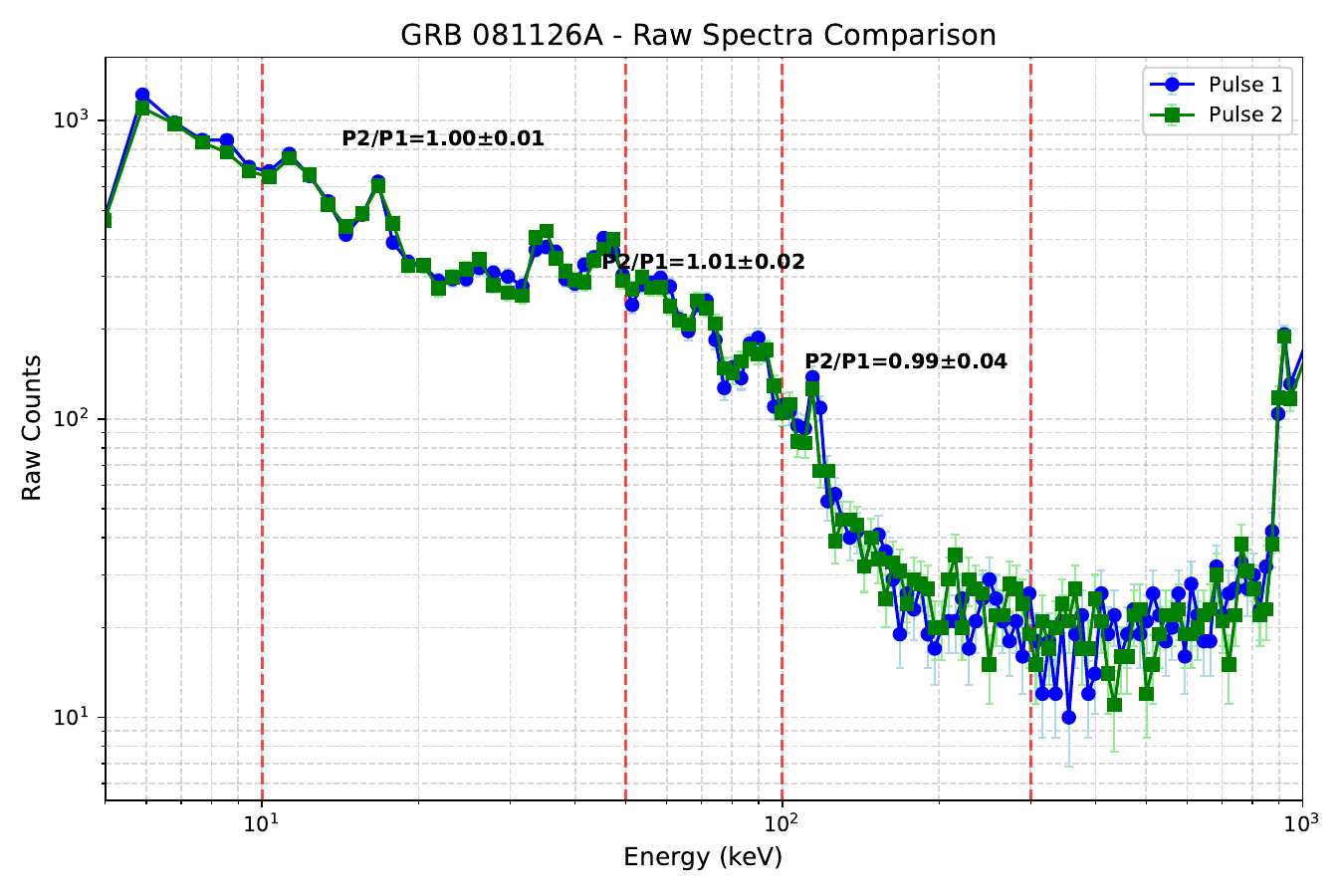}
\includegraphics[width=0.3\textwidth]{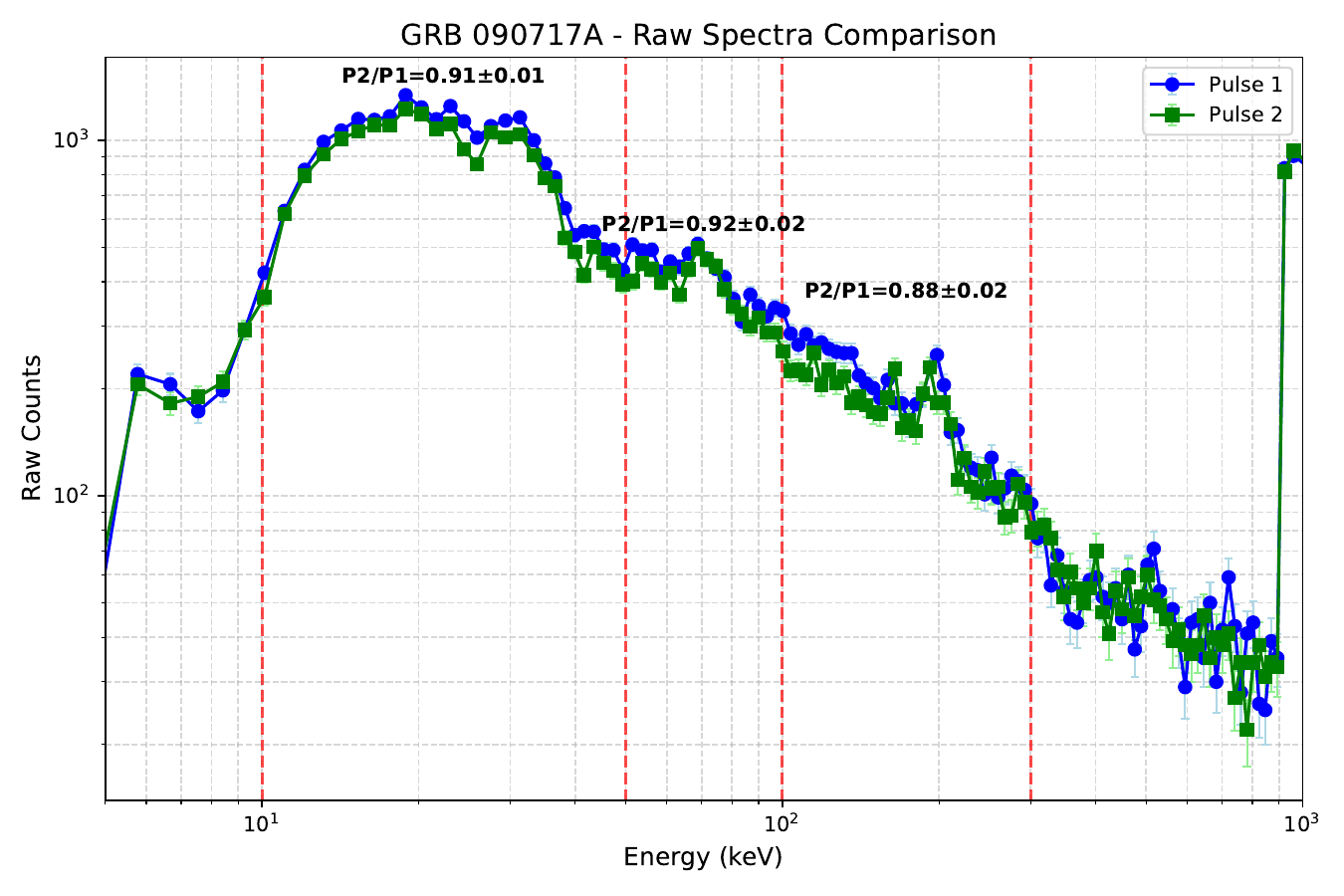}
\\
\includegraphics[width=0.3\textwidth]{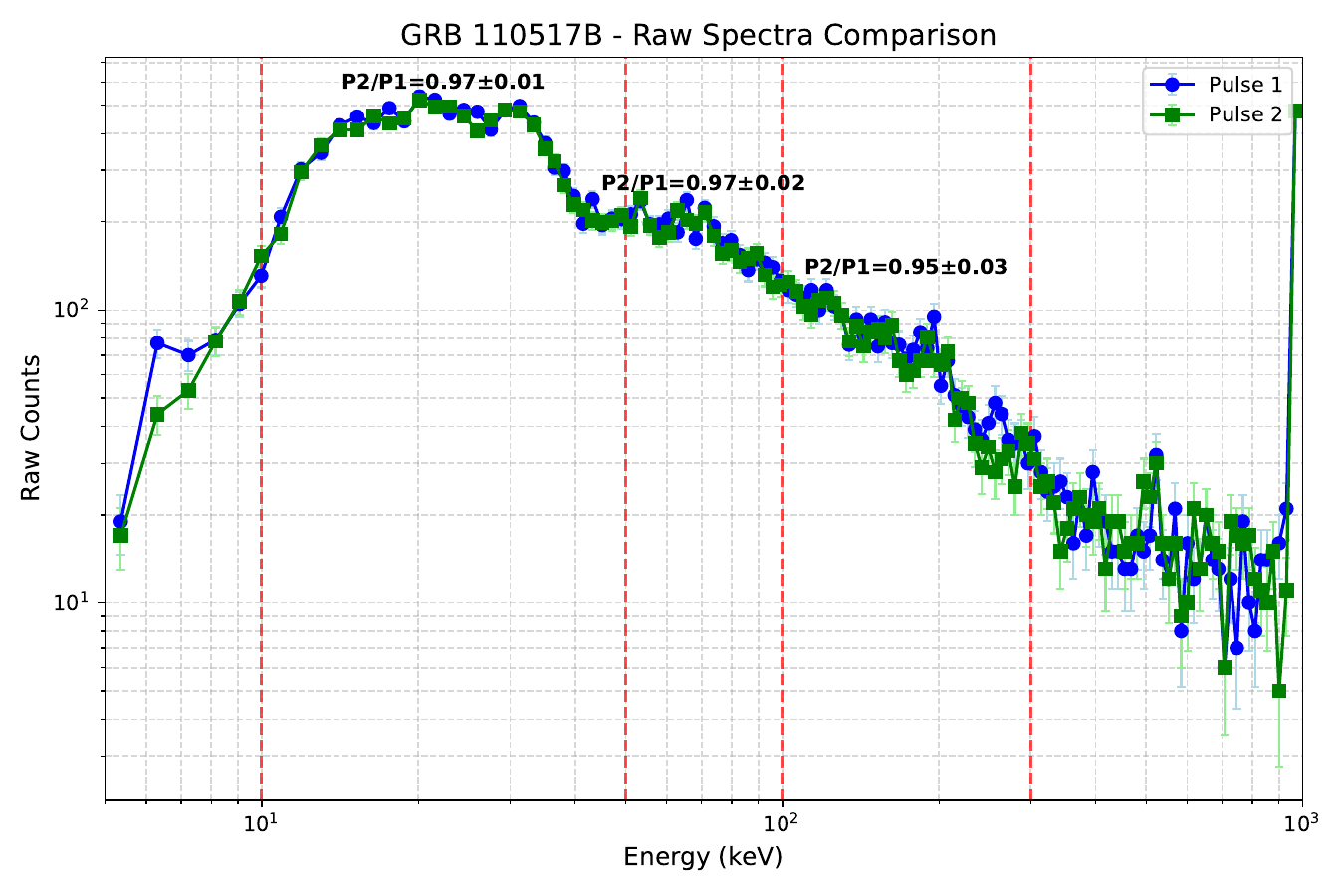}
\includegraphics[width=0.3\textwidth]{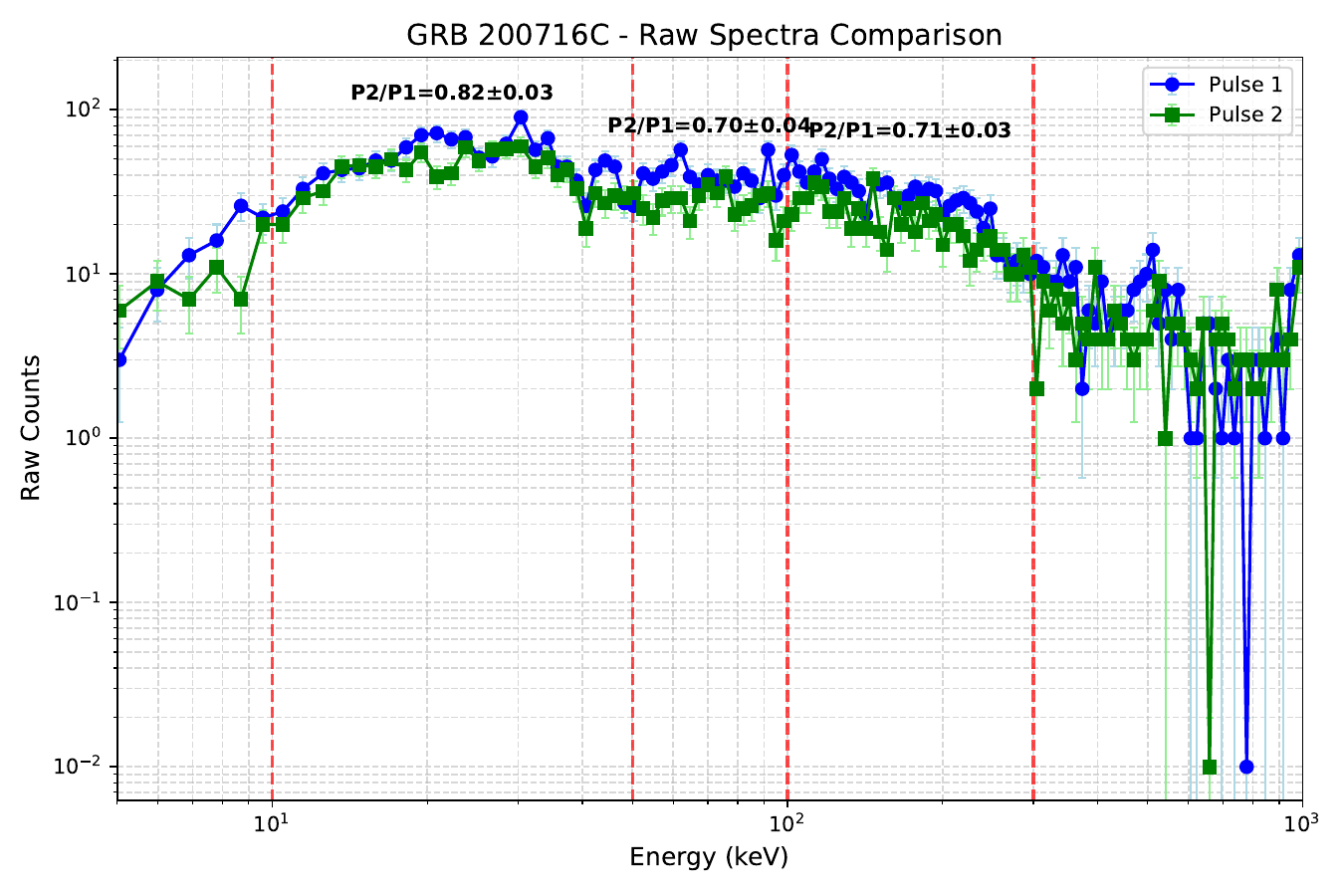}
\includegraphics[width=0.3\textwidth]{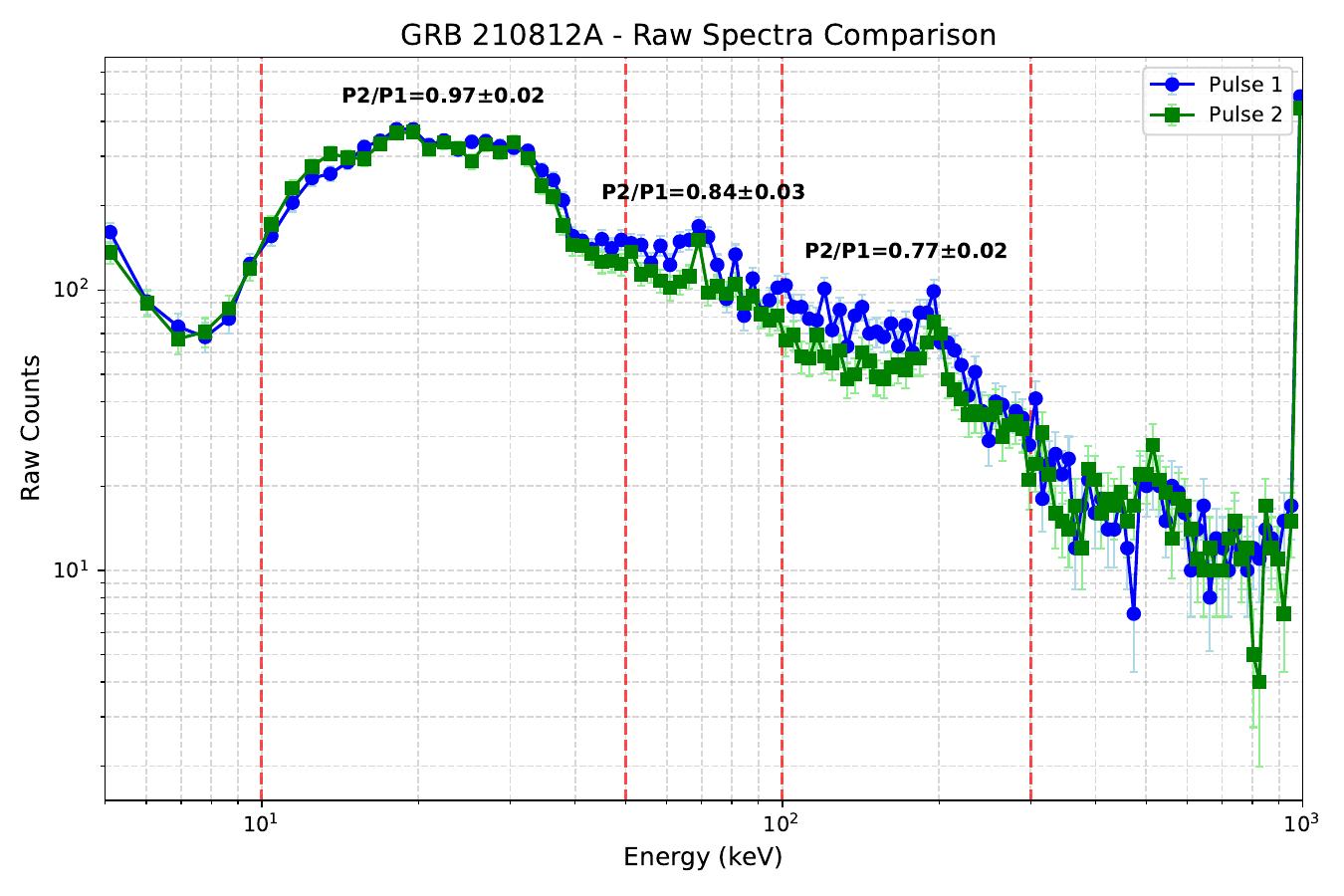}
\caption{Direct comparison of the raw spectra for the two pulses of each GRB. The four red dashed lines from left to right correspond to 10, 50, 100, and 300 KeV, respectively. Here P2/P1 denotes the ratio of the photon counts in the second pulse to those in the first pulse, computed in each energy band, providing a direct measure of the relative spectral consistency between the two pulses.
}
\label{fig 7}
\end{figure}
\subsubsection{Spectral hardness}
Hardness ratio tests have been widely used to support the gravitational lensing hypothesis \citep{2021ApJ...922...77K,2021NatAs...5..560P,2021ApJ...918L..34W,2021ApJ...921L..30V}, based on the assumption that the flux ratio between lensed pulses should not depend on energy\citep{1986ApJ...308L..43P}. In this study, we use the spectral analysis results from the Band function to calculate the HR for each GRB, as shown in Table\ref{table 1}. The HR of each GRB is determined by the ratio of the fluence in the energy range 50-300 keV to the fluence in the energy range 10-50 keV, as defined below:
\begin{equation}
HR = \frac{\int_{50\,\text{keV}}^{300\,\text{keV}} E \cdot N(E)_{\text{Band}} \, dE}{\int_{10\,\text{keV}}^{50\,\text{keV}} E \cdot N(E)_{\text{Band}} \, dE}
\label{eq:HR}
\end{equation}
We take GRB 081126A as an example, and the HRs of the two pulses are consistent: ${\text{HR}}_1 = 5.52 \pm 0.83$ and ${\text{HR}}_2 = 5.05 \pm 0.98$. For GRBs (081126A, 090717A, 110517B, 200716C), the HRs of their two pulses are consistent within the $1\sigma$ confidence level. For GRB 081122A and GRB 210812A, the HRs of their two pulses are consistent within the $2\sigma$ confidence level. In addition to the model-based spectral HR derived from fitted Band parameters, we also compute counts-based HR directly from the observed photon counts in the 10–50 keV and 50–300 keV bands as a model-independent check. The results are summarized in Table \ref{table 1}. For example, GRB 081126A shows consistent HRs for its two pulses (${ \mathrm{HR}}{\mathrm{count},1} = 1.27 \pm 0.13$ and ${ \mathrm{HR}}{\mathrm{count},2} = 1.19 \pm 0.16$). Overall, we find that GRBs 081126A and 110517B are consistent within the $1\sigma$ level, while GRBs 090717A and 200716C are consistent within $2\sigma$ but not within $1\sigma$. For GRBs 081122A and 210812A, the differences exceed the $2\sigma$ level. Since any data fitting process inevitably involves loss of information the rawer the data used for comparison, the better the outcome. Figure~\ref{fig 7} further illustrates these results by showing the raw photon spectral comparison of the two pulses for each burst across three energy bands (10–50, 50–100, and 100–300 keV). From the the ratios of the photon counts in the second pulse to those in the first pulse showed in Figure~\ref{fig 7} we can find that the model-independent comparison yields results that are broadly consistent with the HR$_\mathrm{count}$ analysis but with higher sensitivity. GRBs 081126A and 110517B show close agreement between their two pulses within the $1\sigma$ level, and GRB 090717A exhibits agreement within $2\sigma$. In contrast, GRB 081122A and GRB 210812A differ by more than $2\sigma$, while for GRB 200716C the raw spectra reveal discrepancies exceeding $3\sigma$, stronger than suggested by the HR$_\mathrm{count}$ test.

\subsubsection{Amati Relation}
\begin{figure}
    \centering
    \includegraphics[width=\textwidth]{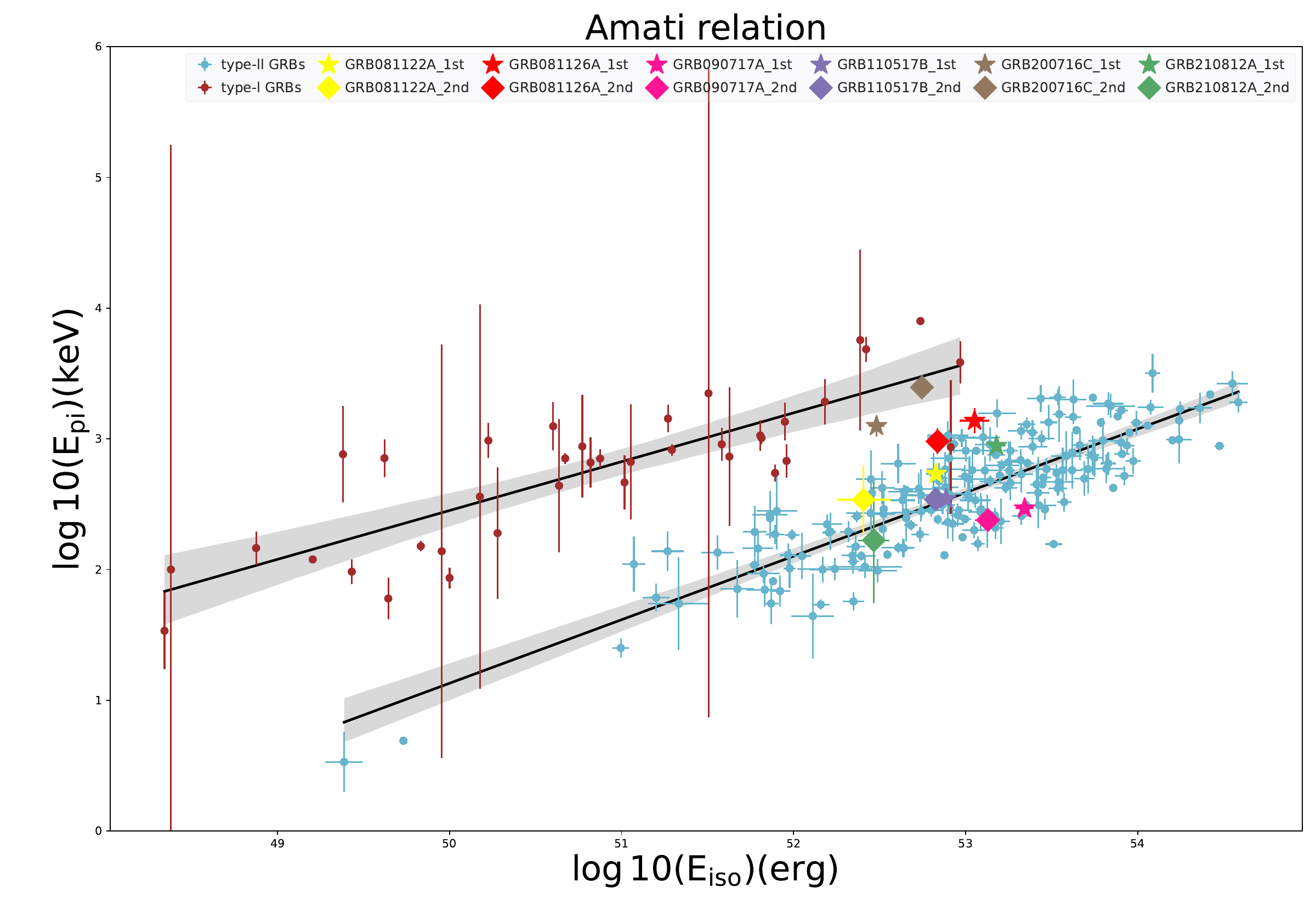}
    \caption{Amati relation. In the figure, maroon circles represent Type I GRBs, and blue circles represent Type II GRBs. The first pulses of the sample GRBs are shown as pentagrams, while the second pulses are shown as diamonds. GRBs with the same color correspond to the same event.}
    \label{fig 8}
\end{figure}

The Amati relation, proposed by \cite{2002A&A...390...81A}, is widely employed for classifying GRBs. In this context, \(E_{p,z}=(1+z)E_p\) represents the rest-frame peak energy, and \(E_{\gamma,iso}\) denotes the isotropic equivalent energy. The expression for \(E_{\gamma,iso}\) is given by:
\begin{equation}
E_{\gamma,iso} = \frac{4\pi d_L^2 kS_{\gamma}}{1+z},
\end{equation}
where \(d_L\) is the luminosity distance, \(S_{\gamma}\) is the fluence in erg \(\text{cm}^{-2}\), and \(k\) is a correction factor used to convert the observed energy range to the GRB rest frame. The correction factor \(k\) is defined as:
\begin{equation}
k = \frac{\int_{\frac{E_1}{1+z}}^{\frac{E_2}{1+z}} EN(E) dE}{\int_{e_1}^{e_2} EN(E) dE}, 
\end{equation}
where \(E_1 = 1\) keV and \(E_2 = 10^4\) keV define the integration limits in the rest frame, and \(e_1 = 8\) keV and \(e_2 = 40\) MeV correspond to the energy range covered by Fermi-GBM in the observer frame.
For GRBs with unknown redshifts, we adopt a typical assumed redshift value of 2. As shown in Figure \ref{fig 8}, the pulses of all the bursts follow the Amati relation and cluster within the same region.

\section{Discussion and Conclusions  \label{section 5}}
We re-evaluated the gravitational lensing candidacy of six GRBs by conducting comprehensive temporal and spectral analyses. Each GRB exhibits two emission phases (pulses) hypothesized to be lensed images of the same parent pulses.

Temporal analysis reveals that the T$_{90}$ durations of GRB 200716C are consistent within 1$\sigma$, while those of other GRBs differ significantly ($>$3$\sigma$). Spectral lags in GRB 200716C exceed 3$\sigma$, whereas other GRBs show consistency within 1$\sigma$. Norris function parameters (pulse width, rise time, decay time) exhibit significant inconsistencies across all bursts.  $\chi^2$ tests indicate that GRB 210812A (p=0.5828) is consistent with originating from the same pulse shape (p $>$ 0.05), while GRB 081122A (p=0.0002), 081126A (p=0.0199), 090717A (p=0.0135), GRB 110517B (p=0.00001) and GRB 200716C (p=0.00001) show significant deviations. For the consistency of the HR$_\mathrm{count}$ values between the two pulses of each GRB, we find that GRBs 081126A and 110517B are consistent within the $1\sigma$ level. GRBs 090717A and 200716C are consistent within $2\sigma$ but not within $1\sigma$. For GRBs 081122A and 210812A, the differences exceed the $2\sigma$ uncertainty. The raw spectra count ratios for GRB 081126A and GRB 110517B are consistent within $1\sigma$, and those for GRB 090717A agree within $2\sigma$. In contrast, GRB 081122A shows discrepancies exceeding $2\sigma$, while GRB 200716C and GRB 210812A exhibit differences at a significance level greater than $3\sigma$. The results of raw spectral counts are roughly consistent with the HR$_\mathrm{count}$ analysis, with only some deviations for GRB 200716C. 

Spectral analysis shows that all time-resolved spectra contain clear thermal components. Four GRBs (081122A, 081126A, 110517B, 200716C) exhibit internally consistent $\alpha$ evolution, while all GRBs (expect GRB 210812A) follow flux-tracking (f.t) $E_p$ evolution. GRB 200716C uniquely displays a "dual-tracking" spectral evolution. The integrated spectral parameters ($\alpha$ and $E_p$) of all GRBs in both pulses are consistent within the 1$\sigma$ confidence level, except for GRB 200716C ($E_p$). For all GRBs, the spectral parameters of their two pulses are consistent within the 2$\sigma$ confidence level. HRs are consistent within 1$\sigma$ for GRB 081126A, 090717A, 110517B, and 200716C, and within 2$\sigma$ for GRB 081122A and 210812A. Raw spectra show that the count ratios of GRB 081126A and GRB 110517B agree within $1\sigma$, and that of GRB 090717A agrees within $2\sigma$. In contrast, the difference between GRB 081122A exceeds $2\sigma$, while the difference between GRB 200716C and GRB 210812A exceeds $3\sigma$ at the significance level. All bursts lie within the same region of the Amati relation. By exploring a redshift range from 0.384 to 5, we observe that the classification of the two pulses remains consistent for each GRB. 

To clarify our judgment process, we summarize here the operational criteria adopted in the various tests and discuss their limitations. First, model-independent assessments—the $\chi^2$ similarity test and the HR$_\mathrm{count}$ test—serve as robust checks that are largely independent of specific pulse models. If either test reveals a statistically significant inconsistency, it is sufficient to reject the lensing interpretation of the burst. It should be noted, however, that the $\chi^2$ similarity test, like any method, has limitations and is sensitive to the choice of time bin and window. We select appropriate time windows and conduct exploratory analyses across a range of time bins (0.01--1.024 s) to identify suitable bin sizes. The choice of time bin is critical: within a consistent time window, changing the bin size can affect the results. Smaller bins may allow noise to dominate the pulse morphology, yielding artificially high $p$-values and unreliable results, while larger bins may smear out pulse structure details, also compromising reliability. Therefore, selecting an intermediate bin size enhances the robustness of the results. Second, the $T_{90}$ measure is designed to approximately preserve the peak flux \citep{1995AAS...186.5301K}. In practice, however, the decaying tails of pulses at lower peak fluxes may blend into the background, potentially underestimating $T_{90}$ for fainter pulses \citep{1996ApJ...463..570K}. $T_{90}$ provides valuable information for comparison, although it also has inherent limitations. Finally, parametric models (e.g., Norris pulse fitting; spectral Band/CPL/BB models) compress complex behaviors into a small set of interpretable parameters, though fitting may obscure genuine differences. We emphasize that similar best-fit spectral parameters do not guarantee identical intrinsic spectra, parametric fits can mask real differences. Therefore, spectral consistency alone is not definitive proof of lensing. To mitigate information loss, we consistently combine parametric comparisons with other model-independent tests. As shown in Figure~\ref{fig 7}, we directly compare the raw photon counts across energy channels, providing a fully model-independent check to avoid potential information loss due to parametric fitting. Each test carries scientific significance, and all results are considered collectively to evaluate the lensing interpretation. We emphasize that any single independent test showing a significant inconsistency (e.g., a light-curve $\chi^2$ test with $p<0.05$, HR$_\mathrm{count}$ inconsistency, or raw spectral mismatch) is sufficient to disfavor the lensing hypothesis for that burst, while the combination of multiple tests provides a comprehensive assessment despite the limitations inherent to each method.

For GRB 081122A, the $\chi^2$ test yields a probability of $p = 0.0002$ ($p < 0.05$) that the pulses originate from different parent distributions. At the same time, the difference in HR$_\mathrm{count}$ exceeds $2\sigma$. The $T_{90}$ durations differ by more than $3\sigma$, with the second pulse—despite its lower peak flux—exhibiting a significantly longer $T_{90}$ than the first. Furthermore, as shown in Figure~\ref{fig2}, the first pulse exhibits two distinct peaks, while the second contains only one. Despite the good spectral agreement, these features, together with the raw spectra comparison, indicate that GRB 081122A is not a case of gravitational millilensing.

A similar analysis of GRB 081126A shows that the $\chi^2$ test yields a probability of $p = 0.0199$ that the pulses originate from the different parent distribution. HR$_\mathrm{count}$ is consistent within $1\sigma$. However, the $T_{90}$ durations differ by more than $3\sigma$, and the Norris parameters (pulse width, rise time, and decay time) also show discrepancies exceeding $3\sigma$. Thus, although the spectral agreement is good and HR$_\mathrm{count}$ is consistent within $1\sigma$, these features suggest that GRB 081126A is not a typical case of gravitational millilensing.

For GRB 090717A, the $\chi^2$ test yields a probability of $p = 0.00001$ that the pulses arise from different parent distributions. HR$_\mathrm{count}$ is consistent within $2\sigma$, while the $T_{90}$ durations and pulse-fitting parameters differ by more than $3\sigma$. Moreover, the spectral parameter evolution exhibits inconsistencies. These features indicate that GRB 090717A is not a typical case of gravitational millilensing.

GRB 110517B exhibits HR$_\mathrm{count}$ consistency similar to that of GRB 081126A, with the $\chi^2$ test giving a probability of $p = 0.00001$ that the pulses originate from different shapes. The $T_{90}$ durations and pulse-fitting parameters differ by more than $3\sigma$. Furthermore, as shown in Figure~\ref{fig2}, both pulses contain two peaks, but their relative intensities differ significantly. Despite the good spectral agreement, these features indicate that GRB 110517B is not a typical case of gravitational millilensing.

For GRB 200716C, the $\chi^2$ test yields a probability of $p = 0.00001$ that the pulses arise from different shapes. HR$_\mathrm{count}$ is consistent within $2\sigma$, and the $T_{90}$ durations agree within $1\sigma$, but the pulse-fitting parameters and spectral lags differ by more than $3\sigma$. Importantly, the raw spectra reveal discrepancies exceeding $3\sigma$, stronger than suggested by HR$_\mathrm{count}$. Therefore, although the spectral agreement is good, the other characteristics indicate that GRB 200716C is not a typical case of gravitational millilensing.

GRB 210812A shows inconsistencies similar to those of GRB 110517B in terms of $T_{90}$ and pulse-fitting parameters. The $\chi^2$ test gives a probability of $p = 0.5828$ ($p > 0.05$) that the pulses originate from the same shape, but HR$_\mathrm{count}$ shows discrepancies at the $2\sigma$ level. The raw spectra confirm these inconsistencies at a comparable level. Despite the good spectral agreement, these features suggest that GRB 210812A is not a typical case of gravitational millilensing. It is noteworthy that, due to the low signal-to-noise ratio of the second pulse, only two time-resolved spectra could be obtained, preventing an analysis of its spectral parameter evolution.

In summary, our study finds no compelling evidence supporting gravitational lensing for GRBs 081122A, 081126A, 090717A, 110517B, 200716C, and 210812A. These findings are consistent with those of \cite{2024MNRAS.529L..83M,2024MNRAS.527L.132M}. However, compared with the results of \cite{2024MNRAS.529L..83M}, we ultimately rule out the possibility that GRB 210812A is a gravitational lens candidate, and this discrepancy arises from the different evaluation criteria adopted by the two studies. \cite{2024MNRAS.529L..83M,2024MNRAS.527L.132M} performed two tests (the $\chi^2$ light-curve similarity test and the hardness similarity test) and thus retained GRB 210812A—with a light-curve difference of $0.89\sigma$ and energy-channel count ratios consistent with the average ratio in the hardness test. In contrast, this study constructs a comprehensive "temporal + spectral" framework that integrates their two tests with a broader set of diagnostics. Moreover, we adopt a conservative criterion: any single test revealing a significant inconsistency (e.g., a light-curve $\chi^2$ test with $p<0.05$, HR$_\mathrm{count}$ inconsistency, or raw spectral mismatch) is sufficient to exclude the lensing hypothesis. 
For GRB 210812A, our comprehensive temporal and spectral diagnostics indeed identify such critical inconsistencies—specifically, the photon-count–based hardness ratio (HR$_\mathrm{count}$) between its two pulses differs beyond the $2\sigma$ confidence level (Table~\ref{table 1}), and direct comparisons of the raw photon spectra across energy bands (10–50 keV, 50–100 keV, 100–300 keV) further confirm discrepancies exceeding the $3\sigma$ significance level (Figure~\ref{fig 7}). In addition, the $T_{90}$ durations of the two pulses differ by more than $3\sigma$, and their Norris function parameters (pulse width, rise time, decay time) also show significant discrepancies. Under this conservative evaluation framework, the combined inconsistencies across temporal and spectral domains are sufficient to disfavor a lensing interpretation, leading to the exclusion of all six candidates, including GRB 210812A. 
This naturally raises the question: what conditions would, in principle, constitute convincing proof of a real gravitational millilensing event in a GRB? Drawing on previous discussions in the literature \citep{1986ApJ...308L..43P,1992ApJ...389L..41M,1994ApJ...432..478N,2024MNRAS.529L..83M}, we summarize a set of practical criteria: 
(i) statistically consistent agreement across multiple, independent temporal and spectral diagnostics; 
(ii) achromatic flux ratios between the putative images, independent of energy band; 
(iii) achromatic repetition of temporal and spectral structures across different energy ranges; and 
(iv) direct confirmation through the detection of multiple images by independent instruments or detectors, whenever observationally feasible. 
Together, these criteria reflect both theoretical expectations and empirical requirements emphasized in earlier works, and serve as a benchmark for future searches. 
GRBs as the most distant point-like astrophysical sources, serve as prime cosmic probes. Gravitational lensing studies of GRBs enable the identification of compact objects (COs), whereas optimizing millilensing observations necessitates high-sensitivity detectors with higher temporal resolution. Such instruments are crucial for unambiguously constraining lens origins and enabling detailed gravitational lens modelling.

\begin{acknowledgments}
We would like to thank the anonymous referee for constructive suggestions to improve the manuscript and acknowledge the use of the public data from the Fermi data archives. This work is supported by the National Natural Science Foundation of China(grant 12163007, 11763009), the Key Laboratory of Colleges and Universities in Yunnan Province for High-energy Astrophysics.
\end{acknowledgments}

\bibliography{paperR1}{}
\bibliographystyle{aasjournalv7}

\end{document}